%% file: main.tex
\documentclass[
 reprint,
 superscriptaddress,
 amsmath,amssymb,
 aps,
 prl,
 longbibliography,
 floatfix
]{revtex4-2}

\usepackage{graphicx}
\usepackage{adjustbox}
\usepackage{pgf}
\usepackage{pgfplots}
\usepackage{tikz}
\usepackage{dcolumn}
\usepackage{bm}
\usepackage{siunitx}
\usepackage{hyperref}
\usepackage{import}
\usepackage{mathtools}
\usepackage{subcaption}

\usetikzlibrary{arrows.meta, positioning, calc, shapes.geometric}
\usepgfplotslibrary{fillbetween} 
\usetikzlibrary{pgfplots.units}
\usepgfplotslibrary{groupplots}
\usepgfplotslibrary{colorbrewer}

\pgfplotsset{compat=1.17}
\newcommand{\FGT}{\mathrm{Fe}_\mathrm{3-x}\mathrm{GaTe}_\mathrm{2}}
\newcommand{\Hdemmeas}{H^\mathrm{meas}}
\newcommand{\omegamat}{\Omega}

\begin{document}

\title{A Fourier-Space Approach to Physics-Informed Magnetization Reconstruction from Nitrogen-Vacancy Measurements}

\author{Alexander Setescak}
\email{alexander.setescak@univie.ac.at}
\affiliation{Faculty of Physics, University of Vienna, 1090 Vienna, Austria}
\affiliation{University of Vienna, Vienna Doctoral School in Physics, 1090 Vienna, Austria}

\author{Florian Bruckner}
\affiliation{Faculty of Physics, University of Vienna, 1090 Vienna, Austria}

\author{Dieter Suess}
\affiliation{Faculty of Physics, University of Vienna, 1090 Vienna, Austria}

\author{Young-Gwan Choi}
\affiliation{Max Planck Institute for the Chemical Physics of Solids, 01187 Dresden, Germany}
\affiliation{Department of Physics, University of Ulsan, Ulsan 44610, Republic of Korea}

\author{Hayden Binger}
\affiliation{Max Planck Institute for the Chemical Physics of Solids, 01187 Dresden, Germany}

\author{Lotte Boer}
\affiliation{Max Planck Institute for the Chemical Physics of Solids, 01187 Dresden, Germany}

\author{Chenhui Zhang}
\affiliation{Department of Electrical and Computer Engineering, National University of Singapore, Singapore 117583, Singapore}

\author{Hyunsoo Yang}
\affiliation{Department of Electrical and Computer Engineering, National University of Singapore, Singapore 117583, Singapore}

\author{Claire Donnelly}
\affiliation{Max Planck Institute for the Chemical Physics of Solids, 01187 Dresden, Germany}

\author{Uri Vool}
\affiliation{Max Planck Institute for the Chemical Physics of Solids, 01187 Dresden, Germany}

\author{Claas Abert}
\affiliation{Faculty of Physics, University of Vienna, 1090 Vienna, Austria}

\date{\today}

\begin{abstract}
Reconstructing magnetization textures from nitrogen-vacancy (NV) magnetometry stray-field measurements is a challenging, fundamentally ill-posed inverse problem, further complicated by the unknown effective distance between sensor and magnetic material. Here we show that incorporating a micromagnetic energy functional directly into the inversion filters out unphysical, high-energy configurations, while Fourier-space upward continuation of the stray field allows us to simultaneously fit the distance. Applied to measurements of the van der Waals ferromagnet
$\FGT$, it recovers an effective distance estimate of approximately $\SI{81}{\nano\metre}$ and
low-energy configurations that reproduce the observed field. More broadly,
embedding physics directly into the reconstruction turns ill-posed magnetic
inverse problems into transparent, interpretable reconstructions, with
applicability well beyond NV magnetometry.
\end{abstract}

\maketitle

\section{\label{sec:intro}Introduction}
\begin{figure*}[t]
    \resizebox{\textwidth}{!}{\includegraphics{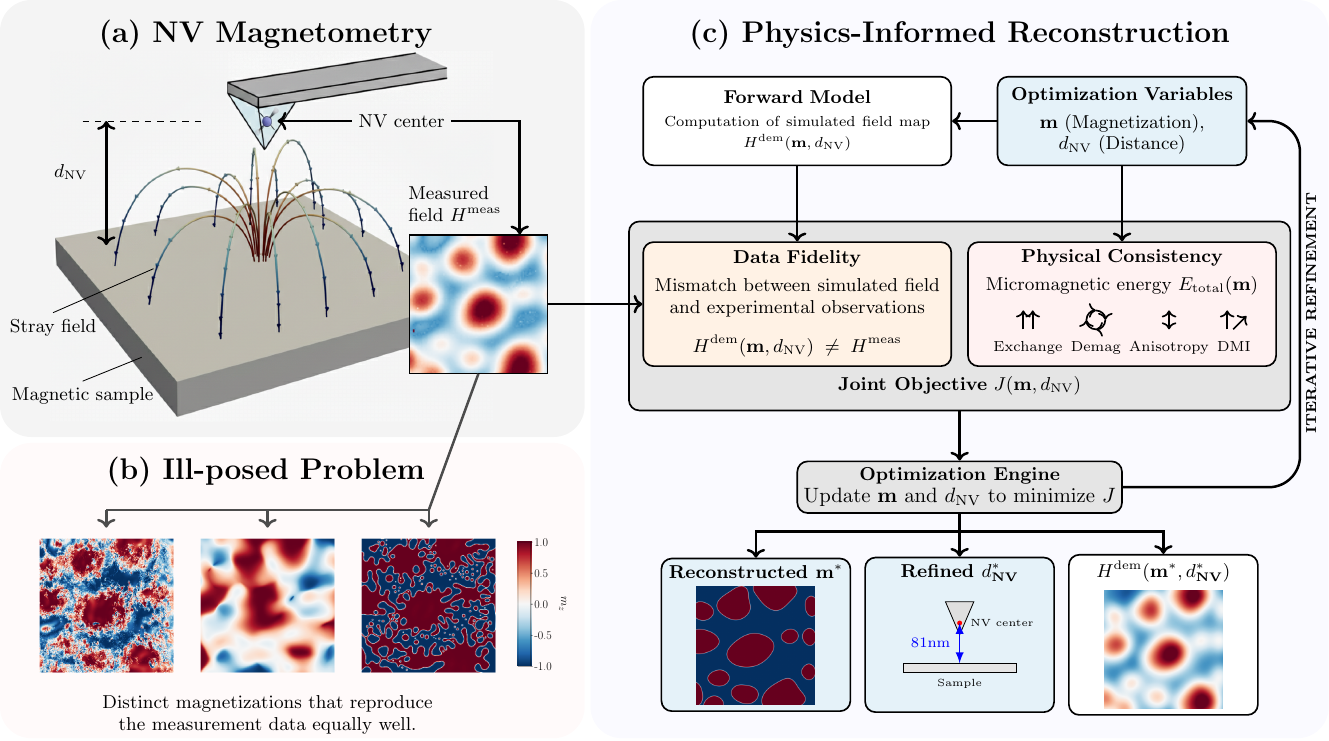}}
    \caption{
        \textbf{Physics-informed reconstruction of magnetization textures from NV magnetometry.} The schematic illustrates the workflow for overcoming the ill-posed nature of stray-field imaging. \textbf{(a)} NV magnetometry measures the stray field projection $H^\mathrm{meas}$ at an unknown effective distance $d_\mathrm{NV}$. \textbf{(b)} Multiple distinct magnetization configurations can produce a nearly identical stray field, making direct inversion unreliable. \textbf{(c)} Our physics-informed framework solves this by coupling a differentiable forward model with a joint objective function. This objective balances data-fidelity with micromagnetic energy. Through iterative gradient-based optimization, the framework simultaneously reconstructs the magnetization $\mathbf{m}$ and refines the effective distance $d_\mathrm{NV}$. This yields the final parameters $\mathbf{m}^\ast$ and $d_\mathrm{NV}^\ast$ as an energetically favorable configuration that remains consistent with the experimental observations.
    }
    \label{fig:nv_and_workflow}
\end{figure*}

Many recent developments in stray-field measurement techniques have enabled the exploration of nanoscale magnetic textures. Among these, nitrogen-vacancy (NV) magnetometry (Fig.~\ref{fig:nv_and_workflow}a) has emerged as a powerful tool: It uses nitrogen-vacancy centers in diamond as atomic-scale quantum sensors. The fluorescence (light emission) from these sensors changes based on their spin state, allowing for the highly sensitive and quantitative measurement of magnetic fields~\cite{casola_probing_2018,xu_recent_2023}. However, inferring the underlying magnetization from measured stray fields constitutes an ill-posed inverse problem~\cite{backus_non-uniqueness_1970,casola_probing_2018,bruckner_solving_2017}: infinitely many magnetization configurations can produce the same stray field (Fig.~\ref{fig:nv_and_workflow}b).
This ambiguity is a hallmark of magnetic inverse problems, also posing significant challenges in the reconstruction of current densities from magnetic field maps~\cite{clement_reconstruction_2021}.
Approaches to solving this inverse problem have evolved across different methodologies. Early work by Hug et al.~\cite{hug_quantitative_1998} established quantitative Magnetic Force Microscopy (qMFM) by determining an instrument calibration function to analytically deconvolve the stray field data. This approach effectively reconstructs the magnetic surface charge equivalent, providing a quantitative map of the field sources at the sample surface~\cite{feng_quantitative_2022}. In contrast, Bruckner et al.~\cite{bruckner_solving_2017} formulated the reconstruction as a variational optimization task. Their approach utilizes the adjoint method and Tikhonov regularization to reconstruct the 3D magnetization by minimizing the field residual via gradient-based optimization. Yao et al.~\cite{yao_universal_2025} additionally incorporate topological charge constraints to successfully reconstruct different magnetic quasiparticles. Simultaneously, the integration of deep learning has further advanced the field. Physically-informed neural networks have proven effective in reconstructing magnetization configurations~\cite{broadway_reconstruction_2025, dubois_untrained_2022}. The network's architecture naturally enforces smoothness, while the field based forward loss enables the recovery of the magnetization. However, when applied to fully three-dimensional spin textures, successful reconstruction remains challenging and requires an initial magnetization guess~\cite{broadway_reconstruction_2025}. 
Recently, we introduced a hybrid strategy~\cite{suess_reconstruction_2025} that utilizes a U-Net to generate an initial guess, followed by micromagnetic relaxation. While this approach is very effective at recovering intrinsic material parameters, it relies on a multistep process that fundamentally separates the data-driven estimation from the final physics-based refinement.
Here, we propose a powerful alternative to existing reconstruction strategies: a self-consistent physics-informed reconstruction framework that embeds micromagnetic physics directly into the optimization (Fig.~\ref{fig:nv_and_workflow}c). Compared to other variational methods that rely on mathematical regularization as a proxy for physics, e.g. using Tikhonov regularization as a qualitative exchange interaction, we replace these heuristic regularizers with quantitative micromagnetic energy terms. This leaves us with a single regularization weight that balances the magnetic stray field residual against the system's total energy, and can be guided by the so-called L-Curve method~\cite{hansen_use_1993,calvetti_tikhonov_2000}. This yields a transparent, end-to-end optimization process that requires no training data and provides quantitative, model-constrained reconstruction of nanoscale magnetic textures along with critical experimental parameters, in particular the \emph{effective distance} $d_\mathrm{NV}$. Throughout this work, $d_\mathrm{NV}$ denotes the total separation between the magnetic material and the NV sensing center, including the NV implantation depth, any surface layers such as sample oxidation, the physical gap between sample and diamond surface, and any other separation between the two~\cite{xu_minimizing_2025}.

\section{\label{sec:forward_model}Forward Model}
Given the magnetization of the sample, our forward model essentially simulates what an NV magnetometer would detect at a given effective distance. It provides us with a differentiable map from a given magnetization configuration $\mathbf{m}$ and the effective distance $d_\mathrm{NV}$ to the expected signal. The required micromagnetic calculations are performed using the open-source, finite-difference Python library \texttt{NeuralMag}~\cite{abert_neuralmag_2025}, configured with its cell-averaged discretization and running on JAX (also verified in the PyTorch-based \texttt{magnum.np}~\cite{bruckner_magnumnp_2023}). Both backends evaluate the stray field through FFT convolution with the demagnetization tensor kernel, resulting in a favorable computational complexity of $\mathcal{O}(N \log N)$. 

All simulations presented in this study focus on an $\SI{800}{\nano\metre} \times \SI{800}{\nano\metre}$ area of a $\FGT$ flake of thickness $\SI{100}{\nano\metre}$. The system is discretized with a cell size of $\Delta_x = \Delta_y = \SI{1}{\nano\metre}$ and $\Delta_z = \SI{100}{\nano\metre}$. The geometry consists of two layers: the bottom layer represents the magnetic material, while the top layer is set to vacuum to calculate stray fields above the sample surface. The measurement itself does not constrain depth variations, and for the chosen parameters the physics-informed regularizer favors a $z$-uniform state. Finer $\Delta_z$ would therefore not recover meaningful new information, and we adopt the single-layer geometry by construction.

The solver computes the stray field averaged over the simulation cell volume ($\SI{1}{\nano\metre} \times \SI{1}{\nano\metre} \times \SI{100}{\nano\metre}$). While lateral spatial averaging is negligible, the vertical averaging over the vacuum layer thickness $\Delta_z$ introduces significant deviations that require explicit correction to enable comparison with measurements performed at a specific effective distance $d_\mathrm{NV}$. Fortunately, in the non-magnetic region $z > z_0$ above the sample surface $z_0$, the magnetic field satisfies Laplace's equation, permitting the derivation of an exact transfer function in 2D Fourier space. Starting with the averaged field $\langle {\mathbf{H}}^{\mathrm{dem}} \rangle_{[z_0, z_0+\Delta_z]}$ in the vacuum layer above the sample, this function recovers the magnetic field at the sample surface and extrapolates it to the effective distance $d_{\mathrm{NV}}$:

\vspace{0.3cm}
\begin{equation}
\tilde{\mathbf{H}}^{\mathrm{dem}}(d_{\mathrm{NV}}) = \langle \tilde{\mathbf{H}}^{\mathrm{dem}} \rangle_{[z_0, z_0+\Delta_z]} \cdot 
\left( \frac{k\Delta_z}{1 - e^{-k\Delta_z}} \right) e^{-k(d_{\mathrm{NV}} - z_0)},
\label{eq:fourier_reconstruction}
\end{equation}
\vspace{0.3cm}

where $k = \sqrt{k_x^2 + k_y^2}$ is the magnitude of the wavevector and $\tilde{\mathbf{H}}^{\mathrm{dem}}$ denotes the stray field in the 2D Fourier domain. The multiplicative factors correspond to de-averaging over the cell thickness of the vacuum layer $\Delta_z$ and upward continuation towards the effective distance $d_{\mathrm{NV}}$, respectively.

Following upward continuation, the vector field $\tilde{\mathbf{H}}^{\mathrm{dem}}(d_{\mathrm{NV}})$ is transformed back to real space via inverse FFT and projected onto the nitrogen-vacancy (NV) center quantization axis $\mathbf{n}_{\mathrm{NV}} = (\sin 54.74^{\circ}, 0, \cos 54.74^{\circ})$, the tetrahedral angle set by the diamond $\langle 111\rangle$ lattice direction. This projection, $H^{\mathrm{dem}} = \mathbf{H}^{\mathrm{dem}} \cdot \mathbf{n}_{\mathrm{NV}}$, yields a final scalar field map that can be directly compared to experimental field measurements.

Notably, our approach decouples the simulation's vertical discretization ($\Delta_z$) from the measurement geometry. This allows for the modeling of thick magnetic samples using coarse discretization while accurately calculating the stray field at arbitrary effective distances. Furthermore, because the entire forward pipeline is implemented in a fully differentiable manner, it enables gradient-based optimization not only for the magnetization field $\mathbf{m}$ but also for auxiliary parameters such as the effective distance $d_{\mathrm{NV}}$. See SM~Sec.~I for the de-averaging and upward-continuation derivation, and SM~Sec.~II.C for the simulation geometry~\cite{supplemental_material}.

\section{Physics-Informed Loss Functional}\label{sec:loss_functional}
To address the inverse problem of inferring a magnetization $\mathbf{m}$ from the measured field $\Hdemmeas$, we formulate an optimization problem by minimizing the loss functional $J(\mathbf{m},d_\mathrm{NV})$:

\begin{equation}
J(\mathbf{m},d_\mathrm{NV}) = \mathcal{L}_{\mathrm{data}}^2(\mathbf{m},d_\mathrm{NV}) + \lambda\,\mathcal{L}_{\mathrm{energy}}(\mathbf{m}).
\label{eq:loss_functional}
\end{equation}

Here $\mathcal{L}_{\mathrm{data}} = \lVert H^{\mathrm{dem}}(\mathbf{m},d_\mathrm{NV}) - \Hdemmeas\rVert / \lVert \Hdemmeas\rVert$ is the relative field error between the modeled and measured stray field, and $\mathcal{L}_{\mathrm{energy}}$ is the total micromagnetic energy $E_{\mathrm{total}}(\mathbf{m})$ normalized to a uniformly out-of-plane magnetized state $\mathbf{m} = \mathbf{e}_z$.

Following the micromagnetic framework of Abert et al.~\cite{abert_micromagnetics_2019}, we define the total energy as the sum of five contributions: the exchange energy ($E^{\mathrm{ex}}$), the demagnetization energy ($E^{\mathrm{dem}}$), the uniaxial anisotropy energy ($E^{\mathrm{aniu}}$), the interfacial Dzyaloshinskii--Moriya interaction (DMI) energy ($E^{\mathrm{dmii}}$), and the Zeeman energy ($E^{\mathrm{Zee}}$) from the applied bias field $\mathbf{H}^{\mathrm{bias}}$. For the anisotropy and DMI formulations, the relevant axis and interface normal directions coincide with the out-of-plane direction $\mathbf{e}_z$. The material parameters used throughout this work are the interfacial DMI constant $D_i \approx \SI{-0.51}{\milli\joule\per\metre\squared}$, saturation magnetization $M_s \approx \SI{53}{\kilo\ampere\per\metre}$, uniaxial anisotropy constant $K_{\mathrm{u}} \approx \SI{0.302}{\mega\joule\per\metre\cubed}$, and exchange stiffness $A \approx \SI{0.70}{\pico\joule\per\metre}$~\cite{zhang_above-room-temperature_2024}. The experimental bias field is $H^\mathrm{bias}\approx\SI{3.2}{\kilo\ampere\per\metre}$. The formulation is not tied to this particular Hamiltonian or geometry: any micromagnetic energy terms and sample topology supported by our differentiable inverse-micromagnetic libraries \texttt{NeuralMag}~\cite{abert_neuralmag_2025} and \texttt{magnum.np}~\cite{bruckner_magnumnp_2023} can be substituted.

The regularization parameter $\lambda > 0$ controls the trade-off between data fidelity and physical stability. We visualize this trade-off with the L-Curve method~\cite{hansen_use_1993}, plotting $\mathcal{L}_{\mathrm{data}}$ against $\mathcal{L}_{\mathrm{energy}}$ for varying $\lambda$ and select a balanced operating point $\lambda_{\mathrm{bal}}$ at the "corner" of the resulting L-Curve.

The loss $J(\mathbf{m},d_\mathrm{NV})$ is minimized by gradient-based optimization, jointly over $\mathbf{m}$ and $d_\mathrm{NV}$, exploiting the differentiability of the forward model. The micromagnetic normalization constraint $|\mathbf{m}|=1$ is enforced via a unit-norm parametrization of the magnetization rather than an additional penalty term. Full expressions for $\mathcal{L}_{\mathrm{data}}$, $\mathcal{L}_{\mathrm{energy}}$, and the five energy contributions, together with the parameter derivation and the optimizer, parametrization, and initialization choices, are given in SM~Sec.~II.

\section{\label{sec:reconstruction}Reconstruction}
\begin{figure}
    \centering
    \resizebox{\columnwidth}{!}{\includegraphics{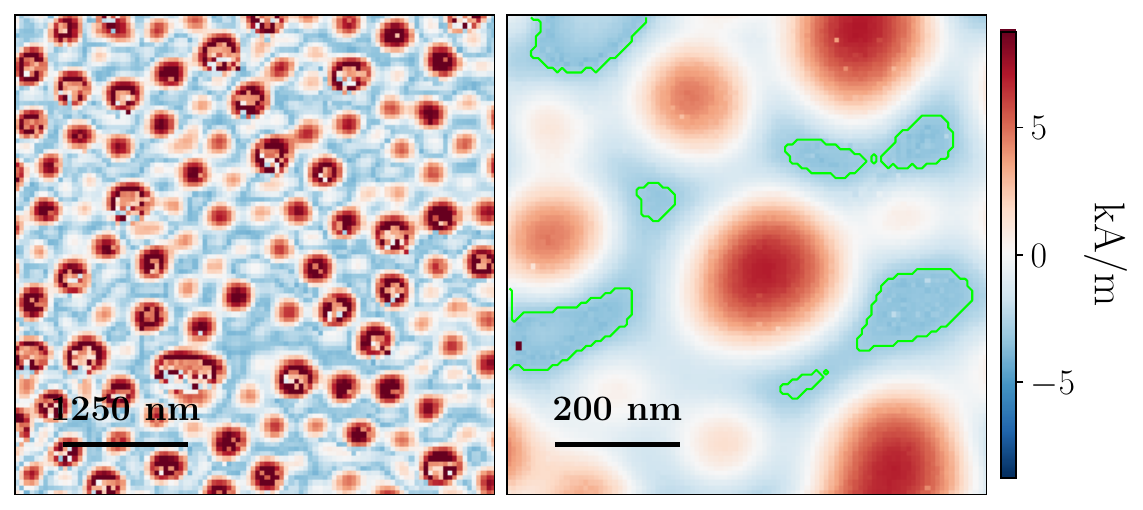}}
    \caption{\textbf{Experimental NV magnetometry measurements of a \SI{100}{\nano\metre} thick $\FGT$ thin film.} Left: large-scale overview ($\SI{5000}{\nano\metre} \times \SI{5000}{\nano\metre}$, $200 \times 200$ pixels). Right: high-resolution scan ($\SI{800}{\nano\metre} \times \SI{800}{\nano\metre}$, $100 \times 100$ pixels) used for the quantitative magnetization reconstruction in this work. The green contour encloses the sign-ambiguity artefact regions, where the opposing projected demagnetizing field $H^{\mathrm{dem}}$ exceeds the bias $H^{\mathrm{bias}}$, identified in practice by a data-driven threshold-and-fill criterion. These regions are excluded from the data-fidelity term $\mathcal{L}_{\mathrm{data}}$. The exact effective distances are unknown.}
    \label{fig:nv_measurements}
\end{figure}

We apply the framework to room-temperature experimental NV magnetometry measurements of $\FGT$ flakes~\cite{zhang_above-room-temperature_2024} (Fig.~\ref{fig:nv_measurements}). Because optically detected magnetic resonance (ODMR) NV magnetometry probes the Zeeman splitting of the NV spin states, it is sensitive only to the field magnitude along the NV axis and cannot resolve its sign~\cite{rondin_magnetometry_2014}. A bias field $H^{\mathrm{bias}}$ is applied to lift this degeneracy, with the experiment reporting the bias-subtracted field $\Hdemmeas = |H^{\mathrm{bias}} + H^{\mathrm{dem}}| - H^{\mathrm{bias}}$ (where $H^{\mathrm{dem}} = \mathbf{H}^{\mathrm{dem}} \cdot \mathbf{n}_{\mathrm{NV}}$). In theory, $\Hdemmeas = H^{\mathrm{dem}}$ everywhere outside localized regions where the opposing demagnetizing field exceeds the bias ($H^{\mathrm{dem}} < -H^{\mathrm{bias}}$). There, the total projected field undergoes a sign inversion, creating ODMR sign-ambiguity artefacts (green contour in Fig.~\ref{fig:nv_measurements}), which we exclude from the data-fidelity term $\mathcal{L}_{\mathrm{data}}$ (see SM~Sec.~III). Masking suppresses the spurious texture induced by these artefacts, even though for this measurement the effect is small.

\begin{figure*}[h!]
\centering
\begin{subfigure}[t]{0.48\textwidth}
\centering
\includegraphics{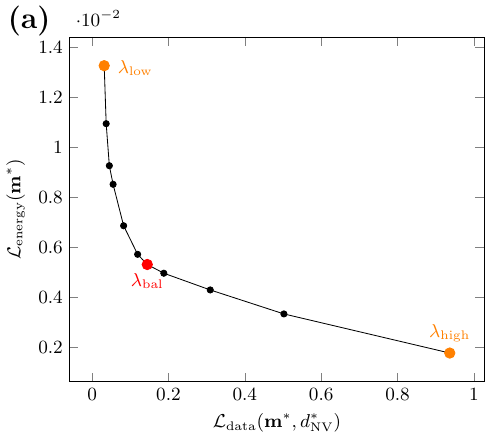}
\end{subfigure}
\hfill
\begin{subfigure}[t]{0.48\textwidth}
\centering
\includegraphics{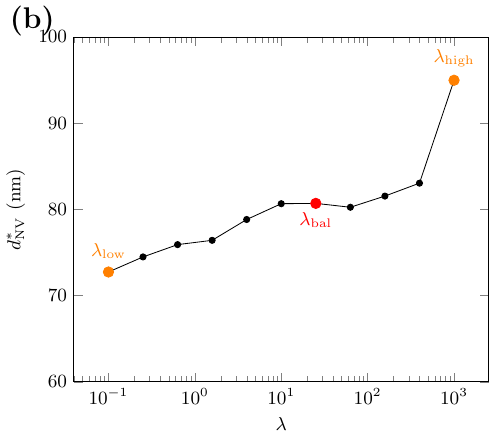}
\end{subfigure}
\caption{\textbf{L-curve analysis for experimental measurement data.} (a) The L-curve illustrates the trade-off between the data-fidelity term $\mathcal{L}_{\mathrm{data}}$ and the normalized regularization energy $\mathcal{L}_{\mathrm{energy}}$. The square marks the balanced operating point $\lambda_{\mathrm{bal}} \approx 25$, where $d_{\mathrm{NV}}^{\ast}$ plateaus (b). (b) Converged effective distance $d_{\mathrm{NV}}^{\ast}$ as a function of the regularization parameter $\lambda$.}
\label{fig:exp_lcurve}
\end{figure*}

The reconstruction jointly optimizes $\mathbf{m}$ and $d_\mathrm{NV}$ using the measured field map $\Hdemmeas$ as input. We repeat this reconstruction independently for a range of regularization strengths $\lambda$. Figure~\ref{fig:exp_lcurve} shows (a) the L-Curve and (b) the corresponding converged effective distances $d_{\mathrm{NV}}^\ast$. We select a balanced operating point $\lambda_{\mathrm{bal}} \approx 25$. We observe that $d_{\mathrm{NV}}^\ast$ rises with $\lambda$ from $\approx \SI{73}{\nano\metre}$ at $\lambda_\mathrm{low} = 0.1$, then plateaus at $\approx \SI{81}{\nano\metre}$ over an order of magnitude in $\lambda$ around $\lambda_\mathrm{bal}$, before climbing to $\approx \SI{95}{\nano\metre}$ at $\lambda_\mathrm{high} = 10^{3}$.

\begin{figure*}
    \centering
    \resizebox{\textwidth}{!}{\includegraphics{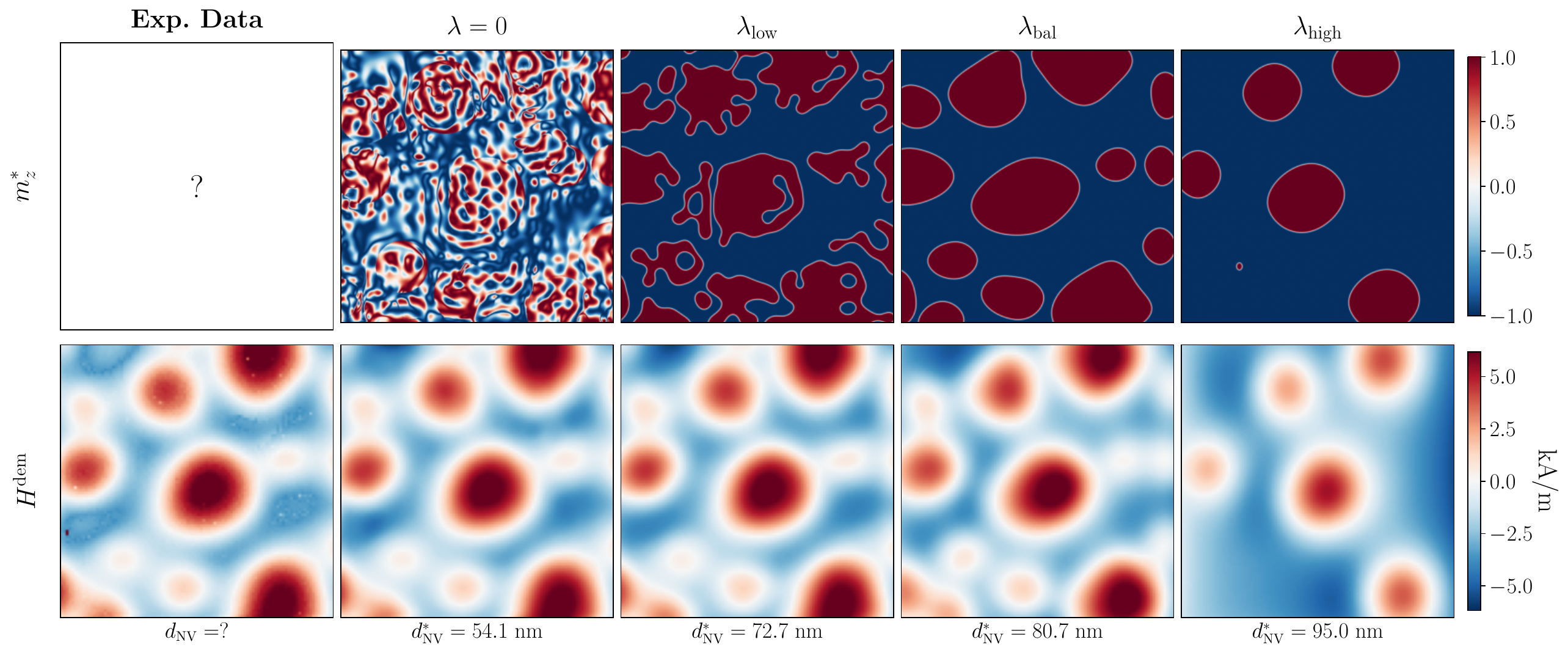}}
    \caption{\textbf{Influence of regularization strength $\lambda$ on reconstructed magnetization and field maps.} Comparison between reconstructed magnetizations $\mathbf{m}^*$ and their corresponding demagnetizing fields $H^{\mathrm{dem}}(\mathbf{m}^\ast,d_\mathrm{NV}^\ast)$ across varying $\lambda$. The first column shows the experimental measurement and the second the unregularized $\lambda = 0$ inversion, which overfits the data into an unphysical, high-energy texture. While higher regularization yields a less diffuse $\mathbf{m}^*$, it results in a poorer fit to the experimental field. At high $\lambda$ values, the field strength is significantly underestimated, indicating a decoupling from the experimental data in favor of the regularization term.
    }
    \label{fig:heatmap_3x3_experimental}  
\end{figure*}

At $\lambda=0$ there is no prior and the inverse problem is under-determined: upward continuation damps high-$k$ stray fields as $e^{-kd_{\mathrm{NV}}}$, so at the large effective distance $d_{\mathrm{NV}}$ the measurement is blind to sub-resolution structure and the reconstruction collapses into high-frequency noise that fits the data perfectly. A weak prior at $\lambda_{\mathrm{low}}=0.1$ suppresses this noise, yet the optimizer still overfits experimental imperfections such as sample inhomogeneities and local variations in saturation magnetization. It can use fine domain patterns to fit the measurement very well at little energy cost, due to the short intrinsic length scale set by the strong out-of-plane anisotropy that keeps domain walls narrow, combined with the DMI that lowers the energy cost of creating them. At smaller $d_{\mathrm{NV}}$ the sensor would resolve these structures directly and prevent this overfitting. At $\lambda_\mathrm{high}=10^{3}$, the energy term dominates and the optimizer favors a strictly physical solution that conflicts with the data. To compensate, $d_\mathrm{NV}$ is artificially increased to blur and dampen the simulated stray field, producing a field noticeably weaker than the measurement. At $\lambda_\mathrm{bal}$, the reconstructed magnetization $\mathbf{m}^\ast$ balances these extremes, minimizing the micromagnetic energy while still reproducing the measured field map fairly well (relative field error $\mathcal{L}_{\mathrm{data}} \approx \SI{14.5}{\percent}$, see Fig.~\ref{fig:heatmap_3x3_experimental}).

A key result is the convergence of the effective distance to $d_{\mathrm{NV}}^\ast \approx \SI{81}{\nano\metre}$, with a plateau-like region in $\lambda$ around $\lambda_\mathrm{bal}$ similar to, but less pronounced than, the plateau observed in reconstructions from synthetic data. Since the contributions to $d_\mathrm{NV}$ are individually unknown in typical experiments, the ability to estimate this composite quantity directly from the measurement data is a significant practical advantage.

For the present setup, we estimate these contributions from the fabrication and setup parameters. The \SI{6}{\kilo\electronvolt} nitrogen implantation places the NV center \num{10} to \SI{20}{\nano\metre} below the diamond surface, the oxidation layer of the uncapped, air-sensitive $\FGT$ surface adds \num{10} to \SI{30}{\nano\metre}, the mechanical contact gap adds a few nanometres, and because the NV can lie anywhere across the \num{200} to \SI{400}{\nano\metre} wide tip, a tip--sample tilt can lift it by a further \num{0} to \SI{40}{\nano\metre}. These contributions sum to a range that contains the extracted $d_{\mathrm{NV}}^\ast \approx \SI{81}{\nano\metre}$, consistent with the \num{26} to \SI{128}{\nano\metre} effective distances reported for comparable scanning-NV probes~\cite{xu_minimizing_2025}. The full budget is given in SM~Sec.~VI~\cite{supplemental_material}.

\section{\label{sec:limitations}Limitations and Outlook}
The first class of limitations is set by the information content of stray-field measurements themselves rather than by the reconstruction. Certain magnetization configurations are inherently invisible to any $\mathbf{H}$-based method. A classic example is a perfectly flux-closed in-plane magnetic vortex, which generates zero external stray field. Additionally, upward continuation acts as a low-pass filter on the stray field~\cite{blakely_potential_1995}, exponentially suppressing high-frequency components as $e^{-k d_{\mathrm{NV}}}$, so any magnetization texture concentrated at high wavevectors $k$ is encoded only weakly in the measurement. In our $\FGT$ thin film, the strong uniaxial out-of-plane anisotropy $K_{\mathrm{u}}$ confines the in-plane magnetization components to narrow domain walls, making their stray-field signatures hard to detect at large distances. The reconstructed wall type and chirality are then determined by the micromagnetic prior rather than by the data. Where the measurement is uninformative, the choice of physics determines the solution.

A second class of limitations stems from the physical model itself: the reconstruction is only as good as the assumed model. Its parameters may be off: a $\pm\SI{20}{\percent}$ uncertainty in the saturation magnetization $M_s$, for instance, propagates into a \SI{10}{\percent} to \SI{14}{\percent} shift in the optimized effective distance $d_\mathrm{NV}^\ast$. It may also be incomplete, neglecting variation of the magnetization along the film thickness $z$, surface roughness, pinning sites, or additional energy terms. Such inhomogeneities are not excluded in principle: once characterized, they can be built into the same differentiable energy and forward model. Two internal consistency checks can flag such effects when they remain unmodeled, although they cannot identify the cause of an inconsistency. The first is the shape of the L-curve itself. When the energy prior and the measurement cannot be simultaneously satisfied, the corner softens and $d_\mathrm{NV}^\ast$ no longer settles onto a clear plateau. Comparison against a synthetic baseline or other datasets makes this easier to recognize (SM~Sec.~IV.B). The second check is dynamic relaxation of the reconstructed state using the Landau-Lifshitz-Gilbert (LLG) equation~\cite{landau_theory_1935,gilbert_phenomenological_2004,bruckner_magnumnp_2023}, which simulates the temporal relaxation of the spin system towards an energy minimum. Because the reconstruction minimizes the joint loss rather than the energy alone, some relaxation is expected even for a correctly specified model. Rapid evolution nevertheless shows that the reconstructed state is not an equilibrium of the assumed model, but it cannot distinguish whether the model is incomplete or whether artifacts and noise imprinted spurious structure on the state (SM~Sec.~IV.C).

Finally, the underlying reconstruction of the magnetization $\mathbf{m}$ from the magnetic field $\mathbf{H}$ remains fundamentally ill-posed. Even with physics constraints, we are not guaranteed to recover the true sample state. Moreover, due to the highly non-convex nature of the micromagnetic energy landscape, gradient-based optimization can converge to different metastable configurations depending on the initialization. Nevertheless, this framework allows us to find significantly better, physically plausible solutions compared to unregularized inversion, and the $(\mathcal{L}_{\mathrm{data}},\mathcal{L}_{\mathrm{energy}})$ plane in which the L-Curve lives provides a physics-grounded criterion for selecting a best guess among these candidate solutions: at equal data fit, prefer the lowest-energy configuration. At equal energy, prefer the better data fit. This allows informed choices about initialization, optimizer, and other hyperparameters (SM~Sec.~II.E and~V), which the data fit alone cannot rank meaningfully.

Looking forward, the same sensitivity to the physical assumptions that bounds the reconstruction is also a source of physical insight. Where the measurement is informative, comparing how different energy models stabilize the reconstructed states at a given data fit can help point to the dominant physical mechanisms. Where it is not, the prior still ensures that the recovered state is a physical solution consistent with the data. This variational approach is also easily extendable: multi-height or multi-orientation NV acquisitions could add complementary constraints that further reduce the non-uniqueness of the inversion in realistic noisy, finite-field-of-view measurements. Moreover, because the framework relies on a differentiable forward model, additional auxiliary experimental parameters (such as relative shifts in multi-distance acquisitions) or even parameters like the saturation magnetization $M_s$ (when $d_\mathrm{NV}$ is independently known) can in principle be co-optimized jointly with the magnetization. Ultimately, this combination of a differentiable forward model and a differentiable micromagnetic energy formulation serves as a template for adapting physics-informed reconstructions to other magnetic imaging modalities.

\section*{Author Contributions}
A.S. conceptualized the specific physics-informed reconstruction framework, implemented the method, performed the synthetic and experimental evaluations, and wrote the original manuscript. This work builds upon the foundational concept of inverse micromagnetics established by F.B. and C.A., who developed the core inverse micromagnetic simulation software \texttt{magnum.np} and \texttt{NeuralMag}. C.A., D.S. and F.B. also directly contributed to the development of the methodology. On the experimental side, the samples were provided by H.Y. and C.Z. Data acquisition was performed by Y.-G.C., H.B., L.B., C.D., and U.V. The NV magnetometry infrastructure is overseen by C.D. and U.V. Additionally, Y.-G.C., U.V. and C.A. supported the application of the method to the experimental data, including the configuration of simulation parameters and the interpretation of the results. All authors engaged in scientific discussions and reviewed the manuscript.

\begin{acknowledgments}
This research was funded in whole, or in part, by the Austrian Science Fund (FWF) 10.55776/PIN1629824, 10.55776/P34671, and 10.55776/PAT3864023 (IMECS). The computational results presented were achieved using the Vienna Scientific Cluster (VSC-5). For the purpose of open access, the author has applied a CC BY public copyright license to any Author Accepted Manuscript version arising from this submission.
\end{acknowledgments}

\bibliography{references}

\clearpage
\onecolumngrid
\begin{center}
    \textbf{\large Supplementary Material for “A Fourier-Space Approach to Physics-Informed Magnetization Reconstruction from Nitrogen-Vacancy Measurements”}
\end{center}

\renewcommand{\thefigure}{S\arabic{figure}}
\renewcommand{\thetable}{S\arabic{table}}
\renewcommand{\theequation}{S\arabic{equation}}
\setcounter{figure}{0}
\setcounter{table}{0}
\setcounter{equation}{0}
\setcounter{section}{0}
\setcounter{secnumdepth}{3}

\section{Derivation of the Fourier-Space Upward Continuation Operator}
\label{sec:fourier_derivation}

This section details the analytical derivation of the transfer function used to upward continue magnetic stray fields from a discretized simulation grid to an arbitrary measurement height (effective distance). This formulation accounts for the finite volume averaging inherent in finite-difference discretizations and allows for continuous differentiability with respect to the effective distance $d_\mathrm{NV}$.

\subsection{Magnetostatic Governing Equations in Fourier Space}

In a source-free region above a magnetic sample, the magnetic field $\mathbf{H}$ satisfies Maxwell's equations for magnetostatics:
\begin{align}
    \nabla \times \mathbf{H} &= 0, \\
    \nabla \cdot \mathbf{H} &= 0.
\end{align}
The curl-free condition implies that the field can be expressed as the gradient of a magnetic scalar potential, $\mathbf{H} = -\nabla \Phi$. Substituting this into the divergence-free condition yields Laplace's equation:
\begin{equation}
    \nabla^2 \Phi = 0.
\end{equation}
We define the geometry such that the magnetic material is confined to the half-space $z < z_0$, and the region $z > z_0$ is free space. To solve this, we apply a two-dimensional Fourier transform over the lateral coordinates $(x, y)$:
\begin{equation}
    \tilde{\Phi}(k_x, k_y, z) = \frac{1}{2\pi}\int_{-\infty}^{\infty} \int_{-\infty}^{\infty} \Phi(x, y, z)\,e^{-i (k_x x + k_y y)}\,\mathrm{d}x\,\mathrm{d}y.
\end{equation}
In Fourier space, Laplace's equation becomes an ordinary differential equation in $z$:
\begin{equation}
    \left( \frac{\partial^2}{\partial z^2} - k^2 \right) \tilde{\Phi}(k, z) = 0,
\end{equation}
where $k = \sqrt{k_x^2 + k_y^2}$ is the magnitude of the in-plane wave vector. For the region $z > z_0$, we require the potential to vanish as $z \to \infty$. Consequently, the physical solution is restricted to the decaying exponential branch:
\begin{equation}
    \tilde{\Phi}(k, z) = \tilde{\Phi}(k, z_0)\, e^{-k(z - z_0)}.
    \label{eq:phi_decay}
\end{equation}
Since the magnetic field components are spatial derivatives of the potential ($\tilde{H}_x = -ik_x\tilde{\Phi}$, $\tilde{H}_y = -ik_y\tilde{\Phi}$, $\tilde{H}_z = k\tilde{\Phi}$), every component of the magnetic field vector $\tilde{\mathbf{H}}$ follows the exact same exponential decay law as the potential:
\begin{equation}
    \tilde{\mathbf{H}}(z) = \tilde{\mathbf{H}}(z_0)\, e^{-k(z - z_0)}.
    \label{eq:field_decay}
\end{equation}

\subsection{Correction for Cell-Averaging}

In finite-difference micromagnetic simulations, the computed stray field $\mathbf{H}^{\mathrm{dem}}$ is typically not a point value at a specific height $z$, but rather a volume-averaged value over a discretization cell of vertical thickness $\Delta_z$.

We define the averaged field $\langle \tilde{\mathbf{H}} \rangle$ over a vacuum layer of thickness $\Delta_z$, extending from the sample surface $z_0$ to $z_0 + \Delta_z$, as follows~\cite{abert_micromagnetics_2019}:
\begin{equation}
    \langle \tilde{\mathbf{H}} \rangle_{[z_0, z_0+\Delta_z]} = \frac{1}{\Delta_z} \int_{z_0}^{z_0 + \Delta_z} \tilde{\mathbf{H}}(z) \, dz.
\end{equation}
Substituting the analytical decay behavior from Eq.~\eqref{eq:field_decay} into this integral yields
\begin{align}
    \langle \tilde{\mathbf{H}} \rangle_{[z_0, z_0+\Delta_z]} &= \frac{\tilde{\mathbf{H}}(z_0)}{\Delta_z} \int_{z_0}^{z_0 + \Delta_z} e^{-k(z - z_0)} \, \mathrm{d}z \\
    &= \frac{\tilde{\mathbf{H}}(z_0)}{\Delta_z} \left[ -\frac{1}{k} e^{-k(z - z_0)} \right]_{z_0}^{z_0 + \Delta_z} \\
    &= \tilde{\mathbf{H}}(z_0) \left( \frac{1 - e^{-k\Delta_z}}{k\Delta_z} \right).
\end{align}
This expression relates the discrete simulation output $\langle \tilde{\mathbf{H}} \rangle$ to the exact field value at the surface $\tilde{\mathbf{H}}(z_0)$. By inverting this relationship, we obtain a de-averaging factor that reconstructs the surface field:
\begin{equation}
    \tilde{\mathbf{H}}(z_0) = \langle \tilde{\mathbf{H}} \rangle_{[z_0, z_0+\Delta_z]} \cdot \left( \frac{k\Delta_z}{1 - e^{-k\Delta_z}} \right).
    \label{eq:de_average}
\end{equation}

\subsection{Upward Continuation}

To determine the magnetic field at an arbitrary height $z$ (where $z > z_0$), we apply the standard upward continuation operator to the reconstructed surface field $\tilde{\mathbf{H}}(z_0)$:
\begin{equation}
    \tilde{\mathbf{H}}(z) = \tilde{\mathbf{H}}(z_0) \cdot e^{-k(z - z_0)}.
    \label{eq:upward_cont}
\end{equation}
Substituting Eq.~\eqref{eq:de_average} into Eq.~\eqref{eq:upward_cont} provides the complete transfer function:
\begin{equation}
\tilde{\mathbf{H}}(z) = \langle \tilde{\mathbf{H}} \rangle_{[z_0, z_0+\Delta_z]} \cdot \left( \frac{k\Delta_z}{1 - e^{-k\Delta_z}} \right) \cdot e^{-k(z - z_0)},
\end{equation}
which is the transfer function used in the main text. The final magnetic field distribution in real space, $\mathbf{H}(z)$, is recovered via the inverse two-dimensional Fourier transform. This formulation ensures that the simulated signal is explicitly differentiable with respect to the measurement height $z$, enabling gradient-based optimization of the effective distance $d_\mathrm{NV}$.

\section{Detailed Simulation Setup}
\label{sec:detailed_setup}
The micromagnetic simulations were conducted using a finite-difference discretization. The results in this work were produced with the open-source library \texttt{NeuralMag}~\cite{abert_neuralmag_2025}, running on JAX and using its cell-averaged discretization. The same forward model was originally prototyped in the functionally equivalent PyTorch-based \texttt{magnum.np}~\cite{bruckner_magnumnp_2023} and yields equivalent results, so this is an implementation choice rather than a physical assumption. To ensure the numerical model accurately reproduces the experimental behavior of the $\FGT$ flake, we selected material parameters based on recent literature and the mesh discretization based on characteristic length scales.

\subsection{Micromagnetic Energy Formulation}
\label{sec:energy_formulation}
Following the micromagnetic framework of Abert et al.~\cite{abert_micromagnetics_2019}, we define the total energy over the domain $\Omega$ as the sum of five contributions: the exchange energy ($E^{\mathrm{ex}}$), the demagnetization energy ($E^{\mathrm{dem}}$), the uniaxial anisotropy energy ($E^{\mathrm{aniu}}$), the interfacial Dzyaloshinskii--Moriya interaction energy ($E^{\mathrm{dmii}}$), and the Zeeman energy ($E^{\mathrm{Zee}}$) due to the applied bias field. These contributions are given by
\begin{equation}
\begin{split}
E_{\mathrm{total}}(\mathbf{m}) =\;
& \underbrace{\int_{\omegamat} A \, |\nabla \mathbf{m}|^2 \, \mathrm{d}V}_{E^{\mathrm{ex}}} \\[0.5em]
& \underbrace{-\frac{\mu_0 M_s}{2}
\int_{\omegamat} \mathbf{m} \cdot \mathbf{H}^{\mathrm{dem}} \, \mathrm{d}V}_{E^{\mathrm{dem}}} \\[0.5em]
& \underbrace{- \int_{\omegamat}
K_u (\mathbf{m} \cdot \mathbf{e}_u)^2 \, \mathrm{d}V}_{E^{\mathrm{aniu}}} \\[0.5em]
& \underbrace{+ \int_{\omegamat} D_i \left[
\mathbf{m} \cdot \nabla (\mathbf{e}_d \cdot \mathbf{m})
- (\nabla \cdot \mathbf{m})(\mathbf{e}_d \cdot \mathbf{m})
\right] \, \mathrm{d}V}_{E^{\mathrm{dmii}}} \\[0.5em]
& \underbrace{- \mu_0 M_s \int_{\omegamat} \mathbf{m} \cdot \mathbf{H}^{\mathrm{bias}} \, \mathrm{d}V}_{E^{\mathrm{Zee}}}.
\end{split}
\label{eq:etotal_components}
\end{equation}
Here, $A$ denotes the exchange stiffness constant, $\mu_0$ the vacuum permeability, $M_s$ the saturation magnetization, $K_u$ the uniaxial anisotropy constant, $D_i$ the interfacial DMI constant, and $\mathbf{H}^{\mathrm{bias}}$ the externally applied bias field. The unit vectors $\mathbf{e}_u$ and $\mathbf{e}_d$ define the uniaxial anisotropy axis and the interface normal direction, respectively. In the present study, both directions coincide with the out-of-plane direction, i.e., $\mathbf{e}_u = \mathbf{e}_d = \mathbf{e}_z$.

\subsection{Material Parameters}
\label{sec:material_parameters}

The material parameters at $T = \SI{295}{\kelvin}$ are adopted from recent work characterizing the same $\FGT$ flake~\cite{zhang_above-room-temperature_2024}. They report an exchange stiffness of $A \approx \SI{0.70}{\pico\joule\per\metre}$, an effective anisotropy of $K_\mathrm{eff} \approx \SI{0.30}{\mega\joule\per\metre\cubed}$, and an interfacial Dzyaloshinskii-Moriya interaction constant of $D_i \approx \SI{-0.51}{\milli\joule\per\metre\squared}$. The sign of $D_i$ follows the convention of Eq.~\eqref{eq:etotal_components} as implemented in \texttt{magnum.np}~\cite{bruckner_magnumnp_2023}. The \texttt{NeuralMag}~\cite{abert_neuralmag_2025} backend used for the reported reconstructions adopts the opposite sign convention for the interfacial-DMI energy, so the same physical chirality is obtained there with $D_i \approx +\SI{0.51}{\milli\joule\per\metre\squared}$.
The saturation magnetization, $M_s$, is derived via a phenomenological domain wall model~\cite{bodenberger_zur_1977}. By matching the experimentally observed domain wall width $w$ and energy density $\delta_w$, $M_s$ is calculated as
\begin{equation}
M_s = \sqrt{\frac{4\pi\beta\,\delta_w}{\mu_0 w}} \approx \SI{53}{\kilo\ampere\per\metre},
\label{eq:Ms_derivation}
\end{equation}
where $\delta_w = \SI{0.22}{\milli\joule\per\metre\squared}$, $w = \SI{0.24}{\micro\metre}$~\cite{zhang_above-room-temperature_2024}, $\mu_0$ is the vacuum permeability, and $\beta$ is a phenomenological fitting parameter that is approximately 0.31 for magnets with high magnetocrystalline anisotropy~\cite{zhang_above-room-temperature_2024}.
The effective anisotropy $K_\mathrm{eff}$ aggregates both crystalline and shape contributions. However, micromagnetic solvers already calculate the demagnetizing field (shape anisotropy). To avoid double-counting this energy, we need to calculate the intrinsic crystalline anisotropy $K_\mathrm{u}$, which is then to be used as the simulation input:
\begin{equation}
K_\mathrm{u} = K_\mathrm{eff} + \frac{1}{2}\mu_0 M_\mathrm{s}^{2}.
\label{eq:ku_calculation}
\end{equation}
This relation isolates $K_\mathrm{u}$ by re-adding the shape anisotropy energy component subtracted in the effective anisotropy formulation~\cite{hubert_magnetic_2014}. This relation is only valid for thin films with out-of-plane uniaxial anisotropy, where the demagnetizing factors are approximately $N_z = 1$ and $N_x = N_y = 0$.

\subsection{Discretization and Geometry}
The spatial discretization was determined by the characteristic magnetic length scales of the system, specifically the exchange length~\cite{hubert_magnetic_2014}, $l_{\mathrm{ex}} = \sqrt{A/K_\mathrm{eff}}$, and the DMI length~\cite{rohart_skyrmion_2013}, $l_{\mathrm{D}} = 2A/|D_i|$. Along the surface normal, the discretization is $\Delta_z = \SI{100}{\nano\metre}$ across two layers: the bottom representing the magnetic material, and the top representing vacuum for stray field calculations. To resolve the magnetic texture accurately and mitigate boundary artifacts, both measurements utilize a lateral discretization with the exact same 100-cell padding on all lateral sides relative to their regions of interest (ROI). For Measurement 1NV, the lateral cell size is $\Delta_x = \Delta_y = \SI{1}{\nano\metre}$, and adding the 100-cell (\SI{100}{\nano\metre}) padding to the $\SI{800}{\nano\metre} \times \SI{800}{\nano\metre}$ ROI yields a $1000 \times 1000 \times 2$ cell simulation grid. For Measurement 2NV, the lateral cell size is $\Delta_x = \Delta_y \approx \SI{0.95}{\nano\metre}$ to cover a larger $\SI{1000}{\nano\metre} \times \SI{1000}{\nano\metre}$ ROI ($1050 \times 1050$ cells) and adding the 100-cell ($\approx\SI{95}{\nano\metre}$) padding results in a grid of $1250 \times 1250 \times 2$ cells.

\subsection{Loss Terms}
\label{sec:field_loss_comp}

To enable a direct pixel-to-pixel comparison between simulation and experiment, the high-resolution simulation data is downsampled to the experimental spatial resolution by block-averaging adjacent cells. Measurement~1NV ($N = 100 \times 100$ pixels, pixel size \SI{8}{\nano\metre}) is matched by averaging $8 \times 8$ blocks from the \SI{1}{\nano\metre} simulation cells. Measurement~2NV ($N = 150 \times 150$ pixels, pixel size $\approx\SI{6.67}{\nano\metre}$) is matched by averaging $7 \times 7$ blocks from the $\approx\SI{0.95}{\nano\metre}$ simulation cells.

The data-fidelity loss is the normalized root-mean-square (RMS) error (NRMSE) between the simulated and measured projected fields over all $N$ pixels:
\begin{equation}
\mathcal{L}_{\mathrm{data}}(\mathbf{m},d_\mathrm{NV}) =
\sqrt{\frac{\displaystyle\frac{1}{N}\sum_{i=1}^{N}\bigl(H^{\mathrm{dem}}_i(\mathbf{m},d_\mathrm{NV}) - H^{\mathrm{meas}}_{i}\bigr)^2}{\displaystyle\frac{1}{N}\sum_{i=1}^{N}\bigl(H^{\mathrm{meas}}_{i}\bigr)^2}}.
\label{eq:ldata}
\end{equation}
A value of $\mathcal{L}_\mathrm{data}=0$ indicates a perfect fit, while $\mathcal{L}_\mathrm{data}=1$ means the RMS mismatch equals the RMS of the measured field (as poor as predicting zero everywhere), so that a value of $0.1$ corresponds to a \SI{10}{\percent} relative error. For masked reconstructions the sums run only over unmasked pixels, and the denominator is computed on the same set. We minimize NMSE (squared form yields cleaner gradients). NRMSE values reported in the figures are obtained by taking the square root of the converged NMSE.

The regularization term is the normalized total micromagnetic energy:
\begin{equation}
\mathcal{L}_{\mathrm{energy}}(\mathbf{m}) = \frac{E_{\mathrm{total}}(\mathbf{m}) - E_{\mathrm{ref}}}{|E_{\mathrm{ref}}|},
\label{eq:lenergy}
\end{equation}
where $E_{\mathrm{ref}}$ is the energy of a uniform out-of-plane state ($\mathbf{m}=\mathbf{e}_z$). Subtraction centers $\mathcal{L}_\mathrm{energy}$ near zero for near-uniform states, and division by $|E_{\mathrm{ref}}|$ makes $\lambda$ dimensionless. While this remains a somewhat arbitrary reference choice, the uniform state represents a lower-energy reference state than the skyrmion lattice configuration ($E_\mathrm{total}(\mathbf{m}) > E_\mathrm{ref}$), meaning we consistently obtain positive values ($\mathcal{L}_\mathrm{energy} > 0$). Under this normalization, $\mathcal{L}_\mathrm{energy} = 0$ corresponds to the uniform out-of-plane state, and a value of $\mathcal{L}_\mathrm{energy} = 1$ is assigned to a hypothetical completely non-magnetic configuration ($\mathbf{m} = \mathbf{0}$, where all micromagnetic energy contributions vanish). The full joint loss minimized is
\begin{equation}
J(\mathbf{m},d_\mathrm{NV}) = \mathcal{L}_{\mathrm{data}}^2(\mathbf{m},d_\mathrm{NV}) + \lambda\,\mathcal{L}_{\mathrm{energy}}(\mathbf{m}),
\label{eq:loss_full}
\end{equation}
consistent with the $\lambda$ values used throughout this supplement.

\subsection{Gradient-Based Minimization}
\label{sec:optimization_supp}

The loss $J(\mathbf{m},d_\mathrm{NV})$ from Eq.~\eqref{eq:loss_full} is minimized using automatic differentiation as provided by the underlying tensor frameworks: either JAX~\cite{bradbury_jax_2021} (used here via \texttt{NeuralMag}~\cite{abert_neuralmag_2025}) or PyTorch~\cite{paszke_pytorch_2019} (used via \texttt{magnum.np}~\cite{bruckner_magnumnp_2023}). Both the magnetization field $\mathbf{m}$ and the effective distance $d_\mathrm{NV}$ are optimized jointly. The unit-norm constraint $|\mathbf{m}|=1$ is enforced by construction (not as a penalty term, which would add a hyperparameter): without it, the minimization of the micromagnetic energy drives the magnitude of $\mathbf{m}$ to grow unbounded (as larger magnitudes can lower certain energy terms without limit), making the regularization meaningless. We use parametrizations that automatically enforce this constraint.

Regardless of the specific update rule, the optimization is initialized from a magnetization guess $\mathbf{m}_0$ (the choice of which is discussed in Sec.~\ref{sec:sens_init}) and an effective-distance guess $d_{\mathrm{NV},0}$ (whose influence is examined in Sec.~\ref{sec:sens_height_init}), and follows the same iterative procedure:
\begin{enumerate}
    \item \textbf{Forward Pass:} Compute the demagnetizing field $H^{\mathrm{dem}}(\mathbf{m}, d_\mathrm{NV})$ and total energy $E_\mathrm{total}(\mathbf{m})$.
    \item \textbf{Loss Evaluation:} Using these quantities evaluate the loss $J(\mathbf{m}, d_\mathrm{NV})$.
    \item \textbf{Backward Pass:} Automatic differentiation computes the gradients of $J$ with respect to the free magnetization variables and $d_\mathrm{NV}$.
    \item \textbf{Update:} An optimizer step is taken to reduce $J$ while keeping $|\mathbf{m}|=1$.
\end{enumerate}
The procedure iterates until a stopping condition (convergence or maximum number of iterations) is met, yielding the optimized quantities $\mathbf{m}^\ast$ and $d_\mathrm{NV}^\ast$.

Two implementation approaches have worked well in our experiments, each paired with a unit-norm parametrization that enforces $|\mathbf{m}|=1$ automatically:

\textbf{(i) Spherical parametrization + L-BFGS.}
L-BFGS is a highly efficient second-order quasi-Newton optimizer. However, because a standard Riemannian formulation of L-BFGS is not readily available, it is paired with a spherical coordinate representation. Each cell is parametrized as $(\theta_i, \phi_i)$ with
\begin{equation}
\mathbf{m}_i = (\sin\theta_i\cos\phi_i,\; \sin\theta_i\sin\phi_i,\; \cos\theta_i),
\label{eq:spherical_param}
\end{equation}
which is unit-norm by construction. The free variables $\theta_i$, $\phi_i$ and $d_\mathrm{NV}$ are updated using L-BFGS (jaxopt~\cite{jaxopt_implicit_2022}, strong-Wolfe line search, history size~100, 2000 iterations for the experimental reconstructions and 1000 for the synthetic validation of Sec.~\ref{sec:synthetic_validation_supp}). While fast and powerful, a key disadvantage of this coordinate-based approach is that a start from $\mathbf{m}=\mathbf{0}$ is impossible. This is the configuration used for the primary reconstructions reported in this work.

\textbf{(ii) Riemannian Adam on the unit sphere.}
Alternatively, we update $\mathbf{m}_i \in S^2$ directly using a Riemannian Adam optimizer. At each step, the Euclidean gradient $\mathbf{g}_i$ is projected onto the tangent space of the sphere at the current state $\mathbf{m}_i$ via $\mathbf{g}_i^\perp = \mathbf{g}_i - (\mathbf{g}_i \cdot \mathbf{m}_i)\mathbf{m}_i$. This projected gradient is used by the Adam optimizer to compute a search direction $\mathbf{v}_i$, which is similarly projected to the tangent space: $\mathbf{v}_i^\perp = \mathbf{v}_i - (\mathbf{v}_i \cdot \mathbf{m}_i)\mathbf{m}_i$. The update is then retracted back onto the sphere by normalization: $\mathbf{m}_i \leftarrow \text{normalize}(\mathbf{m}_i + \mathbf{v}_i^\perp)$. This scheme also accommodates a start from $\mathbf{m}=\mathbf{0}$, even though the zero vector lies outside the unit-sphere manifold. The first step is then not a standard Riemannian update but a Euclidean gradient update whose normalization $\mathbf{v}_i / \|\mathbf{v}_i\|$ projects the iterate onto the unit sphere. All subsequent steps are standard Riemannian updates. This first gradient combines the measurement term with the Zeeman energy, which is linear in $\mathbf{m}$ and therefore the only energy term with a nonzero gradient at $\mathbf{m}=\mathbf{0}$. Since Adam maintains a running memory of the gradient, this initialization direction is seamlessly integrated into subsequent steps, providing a natural initialization trajectory. We use a custom JAX implementation of Riemannian Adam~\cite{becigneul2019riemannianadaptiveoptimizationmethods}, the manifold generalization of Adam~\cite{kingma_adam_2017}, with the standard Adam coefficients ($\beta_1=0.9$, $\beta_2=0.999$, $\epsilon=10^{-8}$) and a step-decay learning rate ($\alpha_0=3.0$, decay factor~$0.3$ every 750~epochs) for 3000~epochs, which we use in Sec.~\ref{sec:sens_init} to compare all initializations on an equal footing. For the PyTorch backend, a Riemannian optimizer is available out of the box in geoopt~\cite{kochurov_geoopt_2020}.

Both approaches yield consistent reconstructions in the well-balanced regularization regime. In our experiments, the spherical + L-BFGS combination is a convenient default when a starting guess with $|\mathbf{m}|=1$ is available: it carries no learning-rate schedule to tune, and the resulting reconstruction lies on the L-curve Pareto front (see Fig.~\ref{fig:init_sensitivity}). When no such starting guess is available, a cold start ($\mathbf{m}=\mathbf{0}$) with Riemannian Adam is an equally valid choice and converges to a comparable reconstruction. The reconstructions reported here are produced with \texttt{NeuralMag}~\cite{abert_neuralmag_2025} on JAX, whose \texttt{jit}/\texttt{vmap} compilation kept the runtime of the $\lambda$-sweeps and $M_s$-sensitivity studies manageable on our hardware. A functionally equivalent PyTorch implementation of the same forward model is available in \texttt{magnum.np}, including a simplified demonstration of magnetization reconstruction from magnetic field data in its inverse-problem documentation~\cite{magnumnp_developers_inverse_2026}.

\section{Analysis of ODMR Sign Ambiguity Artefacts}
\label{sec:odmr_artifacts}

We analyse the NV magnetometry datasets of the $\FGT$ sample to evaluate the impact of Optically Detected Magnetic Resonance (ODMR) sign ambiguity artefacts. Throughout this work, Measurement~2NV serves primarily as a diagnostic and robustness test case for artefact handling and model--data consistency. The quantitative results of the main text are based on Measurement~1NV. NV magnetometry probes the projection of the local magnetic field onto the quantization axis of the nitrogen--vacancy (NV) center. Each NV center defines a fixed crystallographic axis $\mathbf{n}_{\mathrm{NV}}$ along one of the four $\langle 111\rangle$ directions of the diamond lattice. In the present measurements, a single NV orientation is addressed, whose axis lies in the $x$--$z$ plane at the tetrahedral angle $\theta = 54.74^\circ$ relative to the sample normal ($z$-axis), set by the diamond $\langle 111\rangle$ lattice direction:
\begin{equation}
\mathbf{n}_{\mathrm{NV}} = (\sin\theta, 0, \cos\theta).
\end{equation}
The magnetic field component sensed by the NV is therefore the scalar projection of the total magnetic field onto this axis,
\begin{equation}
H_{\parallel} = (\mathbf{H}^{\mathrm{bias}} + \mathbf{H}^{\mathrm{dem}}) \cdot \mathbf{n}_{\mathrm{NV}},
\end{equation}
where $\mathbf{H}^{\mathrm{bias}}$ denotes the applied external bias field and $\mathbf{H}^{\mathrm{dem}}$ the sample stray field.

\begin{figure*}[t]
    \centering
    \begin{adjustbox}{max width=\textwidth, max totalheight=0.78\textheight, keepaspectratio}
      \includegraphics{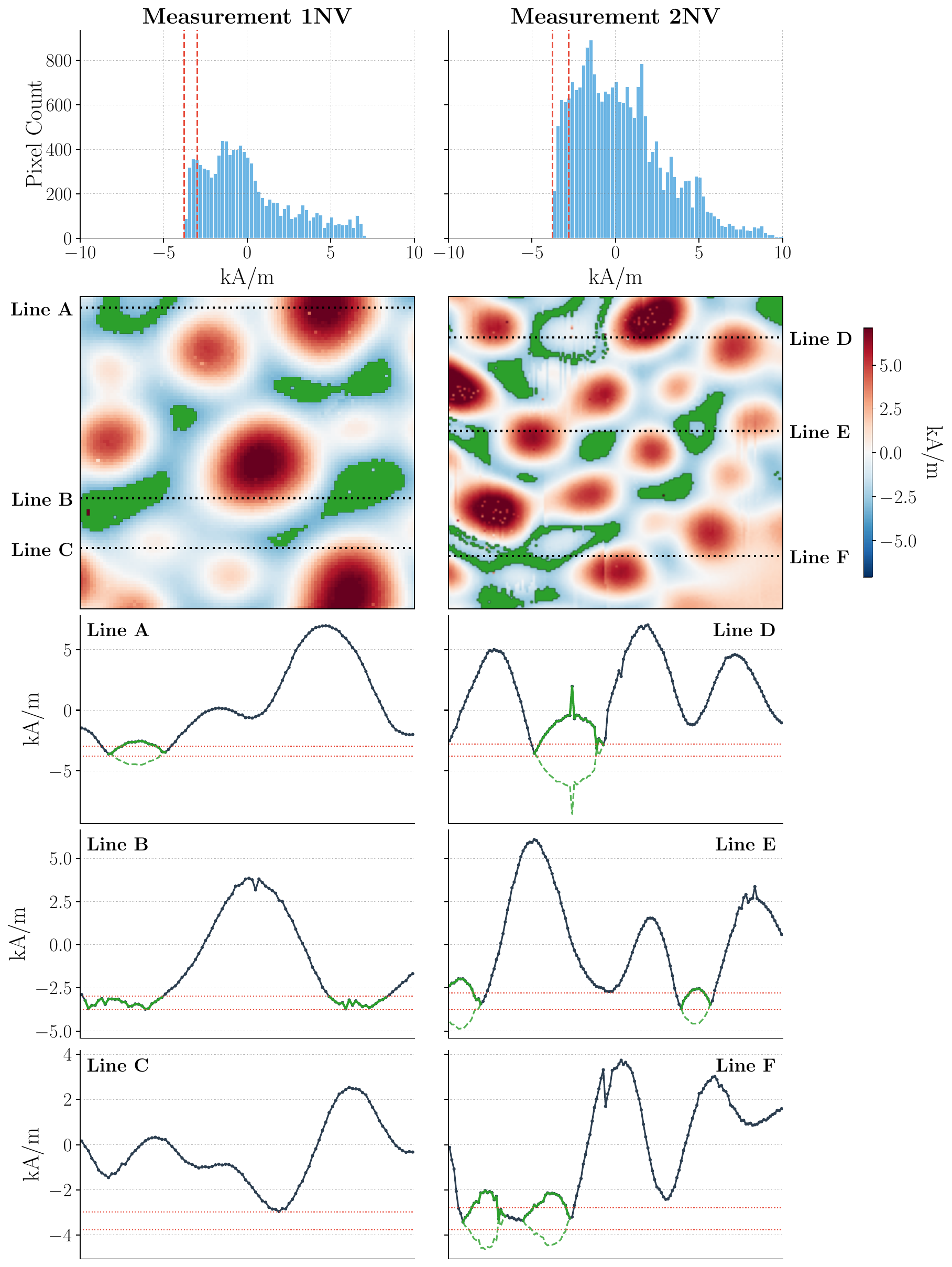}
    \end{adjustbox}
    \caption{\textbf{Analysis of ODMR sign ambiguity artifacts.} Histograms show the magnetic field pixel value distributions for Measurement 1NV and Measurement 2NV. The red dashed lines mark the 0th and 10th percentiles of the field distribution; the green shaded region in the histogram and in the field maps represent the lowest \SI{10}{\percent} of field values. The same percentile thresholds are indicated in the corresponding 1D line cuts (A--F) with red dashed lines. These line cuts illustrate field profiles across selected features, distinguishing artifact-free behavior (Line C), plateau formation (Line B), and folded artifacts (Lines A, D, E, F). In the 1D line plots, solid green segments indicate the artifacts in the measurement and the green dashed lines represent how the magnetic field likely looked in reality.}
    \label{fig:nv_artifact_analysis}
\end{figure*}

ODMR spectroscopy does not directly measure $H_{\parallel}$, but rather the Zeeman splitting of the NV electronic spin resonances. In the presence of a magnetic field component along $\mathbf{n}_{\mathrm{NV}}$, the $m_s=\pm1$ spin states split symmetrically around the zero-field splitting, resulting in a frequency separation
\begin{equation}
\Delta f = 2 \gamma_{\mathrm{NV}} |H_{\parallel}|,
\end{equation}
where $\gamma_{\mathrm{NV}}$ is the NV gyromagnetic ratio~\cite{rondin_magnetometry_2014}. Since ODMR is sensitive only to the magnitude of the frequency splitting, it yields $|H_{\parallel}|$ and is intrinsically insensitive to the sign of the projected magnetic field along the NV axis. A bias field is therefore required for quantitative imaging, as it defines a reference direction and lifts the degeneracy of the spin transitions. In our analysis, we define the reported measurement field $H^{\mathrm{meas}}$, displayed in Fig.~\ref{fig:nv_artifact_analysis}, as the total field magnitude minus the bias contribution:
\begin{equation}
H^{\mathrm{meas}} = |H_{\parallel}| - H^{\mathrm{bias}}.
\end{equation}
Under ideal conditions, where $H_{\parallel} > 0$ across the entire scan, this simplifies to $H^{\mathrm{meas}} = H^{\mathrm{dem}}$. However, if the sample stray field locally opposes and exceeds the applied bias field ($H^{\mathrm{dem}} < -H^{\mathrm{bias}}$), the term $|H_{\parallel}|$ undergoes a sign inversion.

As presented in Fig.~\ref{fig:nv_artifact_analysis}, the pixel value histograms (in \unit{\kilo\ampere\per\metre}) therefore reveal a pronounced asymmetry between positive and negative field values. In Measurement~2NV, positive values extend to approximately \SI{10}{\kilo\ampere\per\metre}, whereas negative values exhibit an abrupt cutoff near \SI{-4}{\kilo\ampere\per\metre}. This cutoff corresponds approximately to the magnitude of the applied bias field (which was $H^{\mathrm{bias}} \approx \SI{4.0}{\kilo\ampere\per\metre}$ for Measurement~2NV, compared to $H^{\mathrm{bias}} \approx \SI{3.2}{\kilo\ampere\per\metre}$ for Measurement~1NV). In practice, $\mathbf{H}^{\mathrm{bias}}$ is not perfectly spatially homogeneous, and small variations in its magnitude across the field of view imply that the precise threshold for sign inversion varies locally. Consequently, instead of marking a single threshold value, we highlight the lowest \SI{10}{\percent} of field values in green. This interval represents a critical regime where the total projected field approaches zero, $|H^{\mathrm{bias}} + H^{\mathrm{dem}}| \approx 0$, and ODMR sign ambiguity artefacts are most likely to occur.

The impact of this ambiguity is clearly visible in the one-dimensional line cuts. Under artifact-free conditions, the stray field exhibits smooth, approximately parabolic local minima. In Measurement~2NV, however, several profiles display deep minima that are rectified into characteristic W-shaped features, as observed in Lines~A, D, E, and F. These profiles indicate regions where $H^{\mathrm{dem}}$ locally opposes and exceeds $H^{\mathrm{bias}}$, causing the measured signal to fold back as the true field crosses zero. In contrast, Measurement~1NV was acquired at a larger effective NV--sample distance, resulting in smaller stray-field amplitudes. As illustrated by Line~B, the corresponding field profiles form plateaus rather than sharp inversions, indicating that the total projected field approaches but does not significantly cross the sign-reversal threshold.

This comparison also enables the validation of subtler magnetic features. For example, the smaller magnetic field texture intersected by Line~C is confirmed to be a genuine feature rather than an artefact, as its local minimum remains well separated from the artifact-prone signal range. Given the minimal presence of critical ODMR sign ambiguity artefacts, Measurement~1NV was selected for the magnetization reconstruction.

\subsection{Impact on Reconstructed Magnetization and Masking Strategy}
\label{sec:masking}

\begin{figure*}
    \centering
    \begin{subfigure}{\textwidth}
        \centering
        \resizebox{\textwidth}{!}{\includegraphics{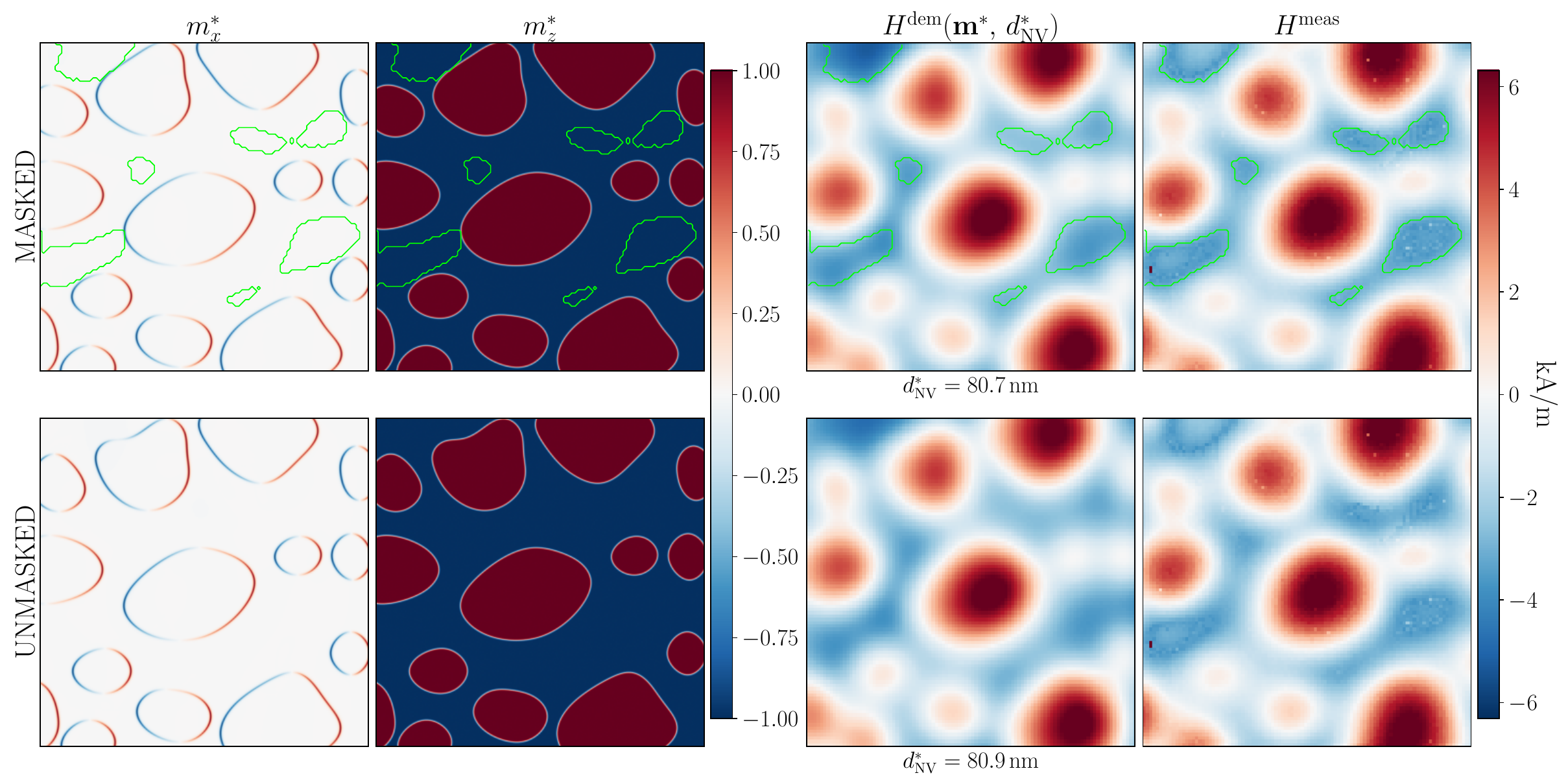}}
        \caption{Measurement~1NV ($\SI{800}{\nano\metre} \times \SI{800}{\nano\metre}$), $\lambda_{\mathrm{bal}} \approx 25$. The masked and unmasked results are nearly identical. A subtle magnetisation distortion is visible in the top-right corner of the unmasked reconstruction, where the optimizer accommodates inconsistent folded-signal pixels.}
        \label{fig:1nv_full}
    \end{subfigure}

    \vspace{1em}

    \begin{subfigure}{\textwidth}
        \centering
        \resizebox{\textwidth}{!}{\includegraphics{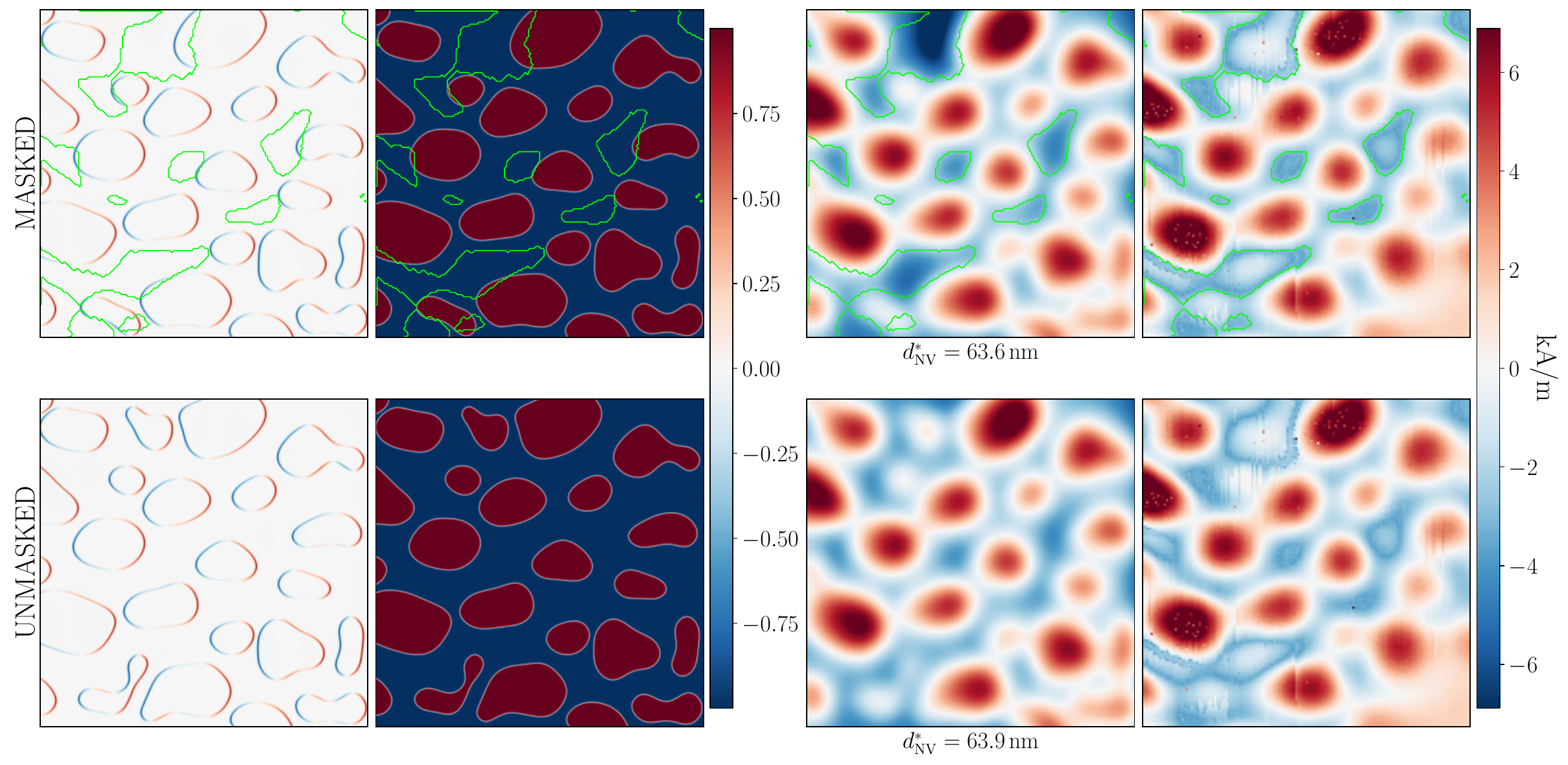}}
        \caption{Measurement~2NV ($\SI{1000}{\nano\metre} \times \SI{1000}{\nano\metre}$), $\lambda_{\mathrm{bal}} \approx 25$. In the unmasked reconstruction, W-shaped rectification artefacts (see Lines~D--F in Fig.~\ref{fig:nv_artifact_analysis}) introduce additional or distorted magnetic textures not present in the masked result.}
        \label{fig:2nv_full}
    \end{subfigure}

    \caption{\textbf{Reconstructions for both experimental NV scan areas.}
    Each panel shows two rows: masked reconstruction (top) and unmasked reconstruction (bottom).
    Columns: reconstructed in-plane component $m_x^\ast$, out-of-plane component $m_z^\ast$,
    simulated stray field $H^{\mathrm{dem}}(\mathbf{m}^\ast, d_{\mathrm{NV}}^\ast)$, and measured field $H^{\mathrm{meas}}$.
    The green contour in the masked row delineates the excluded artefact-prone pixels.
    The inferred effective distance $d_{\rm NV}^\ast$ is indicated below the simulated field column for each row and is not significantly affected by masking.}
    \label{fig:nv_full}
\end{figure*}

The W-shaped rectification artefacts identified above translate directly into spurious features in the reconstructed magnetization. Because the forward model expects a physically consistent stray-field signature, pixels where the measured field has been folded back by the sign ambiguity present an inconsistent target that the optimizer cannot satisfy. In the Measurement~2NV dataset, the pronounced rectification features visible in Lines~D, E, and F introduce false magnetic texture in the reconstructed $m_x^\ast$ and $m_z^\ast$ components (Fig.~\ref{fig:2nv_full}): the unmasked reconstruction shows fine-scale patterns with no physical origin, consistent with the mismatch between the folded measurement and the model-predicted field. The masked reconstruction, which excludes the artefact-prone pixels from the data-fidelity term, suppresses these spurious features and recovers a smoother, more physically plausible texture.

The artefact-prone pixels are identified by a data-driven threshold-and-fill criterion that extends the percentile analysis above to also cover the enclosed, fully folded areas, and the resulting excluded region is outlined by the green contour in the masked reconstruction rows. For Measurement~1NV, where ODMR sign-ambiguity artefacts are mild, masking shifts the inferred $d_\mathrm{NV}^\ast$ negligibly (Fig.~\ref{fig:1nv_full}). For Measurement~2NV, where these artefacts are far more pronounced, masking leaves $d_\mathrm{NV}^\ast$ at the operating point essentially unchanged (\SI{63.6}{\nano\metre} masked vs.\ \SI{63.9}{\nano\metre} unmasked at $\lambda_\mathrm{bal} = 25$). The effect of masking on the stability of the full $d_\mathrm{NV}^\ast(\lambda)$ trace is analysed in Sec.~\ref{sec:model_data_consistency}.

\section{Validation and Consistency Checks}
\label{sec:validation_diagnostics}

\subsection{Validation with Synthetic Data}
\label{sec:synthetic_validation_supp}

We validate the joint reconstruction framework under controlled conditions using synthetic NV measurements generated from a ground-truth magnetization $\mathbf{m}^\mathrm{ref}$ at a reference mean effective distance $d_\mathrm{NV}^\mathrm{ref} = \SI{80}{\nano\metre}$. The reference configuration $\mathbf{m}^\mathrm{ref}$ is generated by taking a reconstruction of the experimental Measurement~1NV (at $\lambda = 10$) and subjecting it to micromagnetic relaxation under the Landau-Lifshitz-Gilbert (LLG) equation (see Sec.~\ref{sec:llg_stability} for details). 

To test the robustness of the effective distance inference, we compare a baseline uniform-distance case against two cases introducing lateral fluctuations in the effective distance (RMS amplitudes $\delta d_\mathrm{amp} = \SI{3}{\nano\metre}$ and $\SI{6}{\nano\metre}$ with a correlation length of $\SI{80}{\nano\metre}$), e.g., arising from sample surface roughness.

\begin{figure*}[!ht]
    \centering
    \resizebox{\textwidth}{!}{\includegraphics{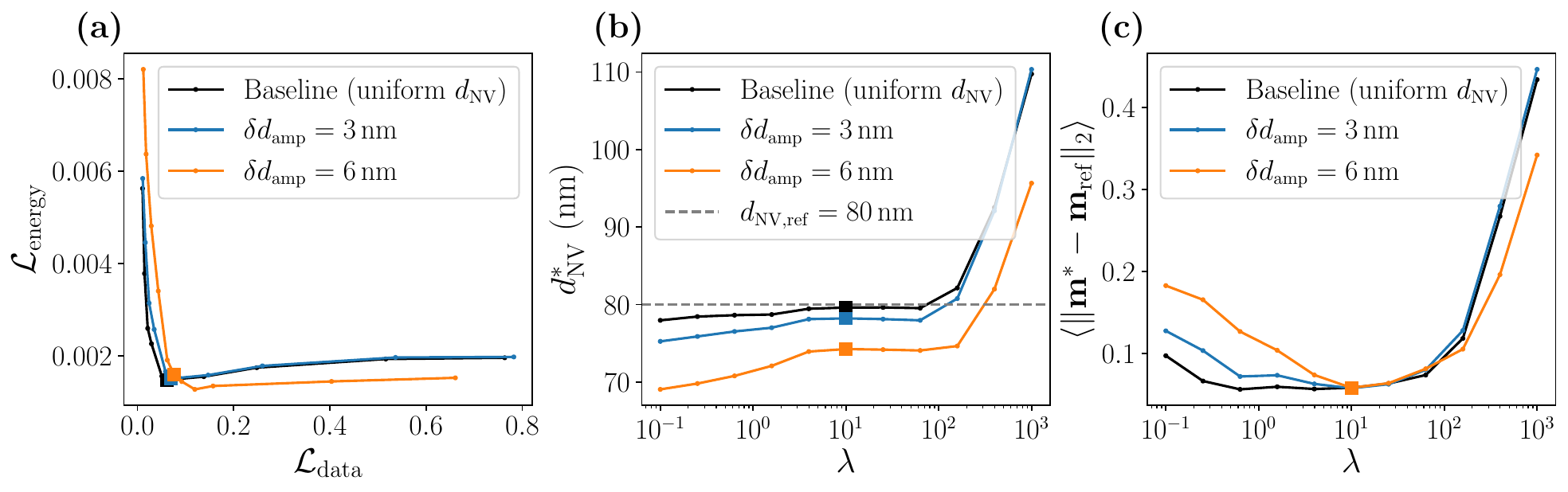}}
    \caption{\textbf{L-curve analysis for synthetic measurement data across three $d_\mathrm{NV}$ scenarios.}
    (a) L-curve (data-fidelity loss vs.\ regularization energy) for the baseline (uniform $d_\mathrm{NV}$) and two rough-surface cases ($\delta d_\mathrm{amp} = \SI{3}{\nano\metre}$ and $\SI{6}{\nano\metre}$) with distance fluctuations. Squares mark $\lambda_\mathrm{bal} \approx 10$.
    (b) Converged effective distance $d_{\mathrm{NV}}^{\ast}$ as a function of $\lambda$. The dashed line marks the ground-truth reference $d_\mathrm{NV}^\mathrm{ref} = \SI{80}{\nano\metre}$.
    (c) Mean magnetization error $\langle\|\mathbf{m}^\ast - \mathbf{m}^\mathrm{ref}\|_2\rangle$ vs.\ $\lambda$. At low $\lambda$ the three scenarios diverge. At $\lambda_\mathrm{bal} \approx 10$ they converge to a common error of ${\approx}0.058$, the minimum for the two roughness scenarios.}
    \label{fig:lcurve_synthetic}
\end{figure*}

To simulate these measurements, the stray fields are computed at discrete vertical slices in Fourier space, interpolated pixel-wise in real space at the local coordinates $d(x,y) = d_\mathrm{NV}^\mathrm{ref} + \delta d(x,y)$, and block-averaged to match the experimental spatial resolution of \SI{8}{\nano\metre}. Starting from an initial guess $d_\mathrm{NV,0}=\SI{80}{\nano\metre}$ and an informed initialization of the magnetization (Sec.~\ref{sec:sens_init}), we jointly optimize $\mathbf{m}$ and the uniform effective distance $d_\mathrm{NV}$. An L-curve sweep determines the balanced regularization parameter $\lambda_\mathrm{bal}\approx 10$, which marks the corner of the L-curve and, for the two roughness scenarios, the minimum of the mean magnetization error $\langle\|\mathbf{m}^\ast - \mathbf{m}^\mathrm{ref}\|_2\rangle$ (Fig.~\ref{fig:lcurve_synthetic}).

\begin{figure*}[!ht]
    \centering
    \resizebox{\textwidth}{!}{\includegraphics{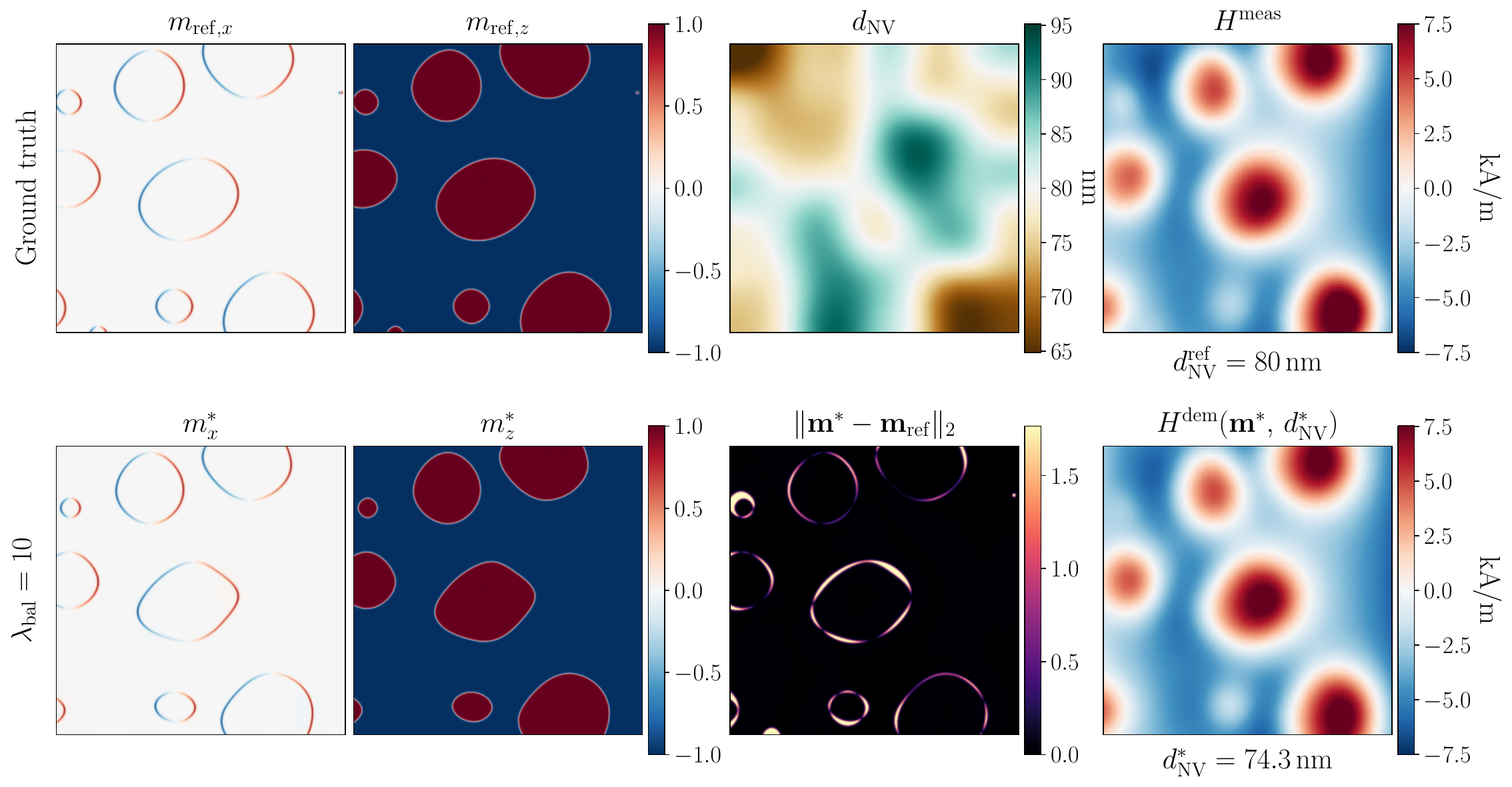}}
    \caption{\textbf{Synthetic reconstruction summary for the $\delta d_\mathrm{amp} = \SI{6}{\nano\metre}$ roughness scenario at $\lambda_\mathrm{bal} = 10$.}
    \emph{Top row:} Ground-truth reference magnetization components $m^\mathrm{ref}_x$ and $m^\mathrm{ref}_z$; effective distance map $d_\mathrm{NV}$ illustrating the \SI{6}{\nano\metre}-amplitude roughness; and the simulated NV projected field $H^\mathrm{meas}$.
    \emph{Bottom row:} Reconstructed components $m^\ast_x$ and $m^\ast_z$; cell-wise L2 error $\|\mathbf{m}^\ast - \mathbf{m}^\mathrm{ref}\|_2$, largest at domain walls; and the model-predicted projected field $H^\mathrm{dem}(\mathbf{m}^\ast, d^\ast_\mathrm{NV})$, which closely reproduces $H^\mathrm{meas}$.}
    \label{fig:synthetic_summary}
\end{figure*}

At $\lambda_\mathrm{bal}$, the mean magnetization error converges to ${\approx}0.058$ for all three scenarios (Fig.~\ref{fig:lcurve_synthetic}(c)), while the inferred effective distance shifts systematically below the ground truth, from \SI{79.6}{\nano\metre} at baseline to \SI{78.2}{\nano\metre} and \SI{74.3}{\nano\metre} at $\delta d_\mathrm{amp} = \SI{3}{\nano\metre}$ and $\SI{6}{\nano\metre}$ (Fig.~\ref{fig:lcurve_synthetic}(b)). This downward bias is a geometric consequence of the exponential field decay: closer regions of a rough surface dominate the sensed amplitude, so the best uniform effective distance fit settles below the mean. The reconstructed magnetization (Fig.~\ref{fig:synthetic_summary}, shown for the \SI{6}{\nano\metre} case) reproduces the domain-wall profile correctly because the wall width and chirality are pinned by the energy regularizer, but the exact domain shape and position are not recovered faithfully: at an effective distance of \SI{80}{\nano\metre} with \SI{8}{\nano\metre} measurement pixels the stray field is largely insensitive to exact structure.

\subsection{Model Accuracy Across Datasets}
\label{sec:model_data_consistency}

\begin{figure*}
    \centering
    \resizebox{\textwidth}{!}{\includegraphics{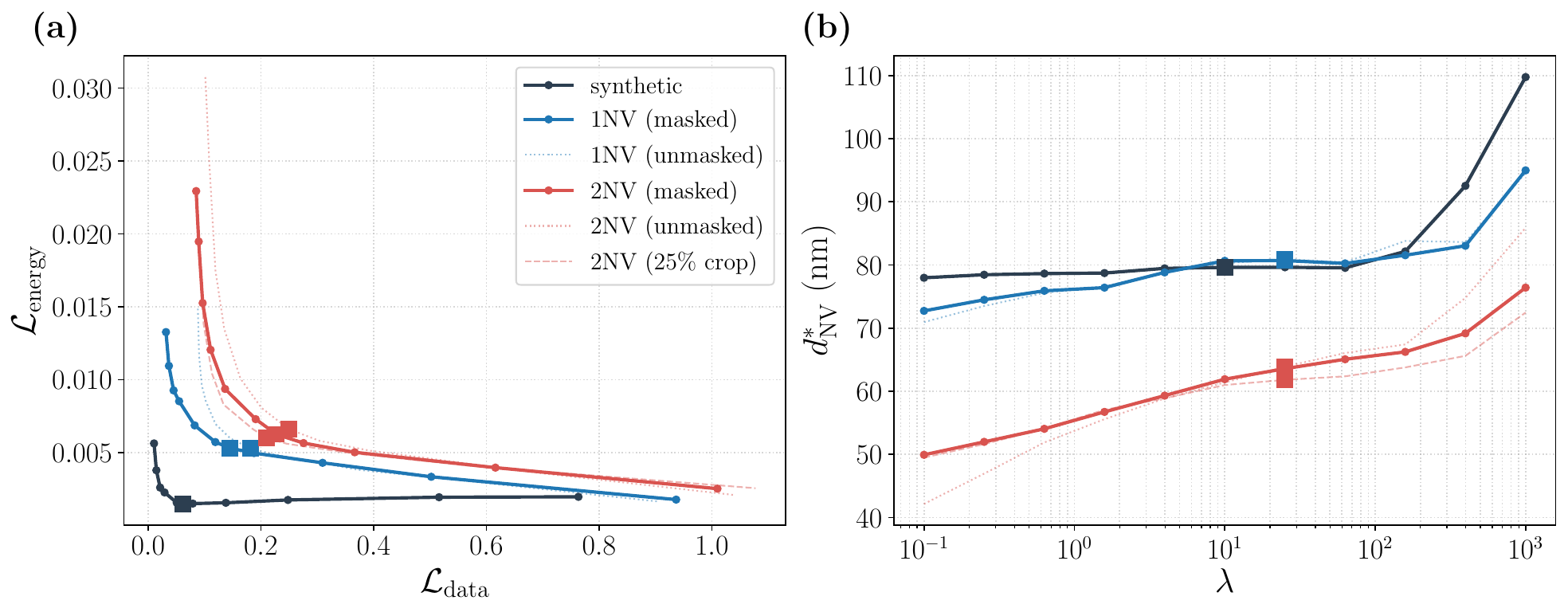}}
    \caption{\textbf{L-curve and effective distance convergence across all three datasets.}
    (a) L-curve (data-fidelity loss vs.\ total energy) for the synthetic baseline, masked/unmasked variants of Measurement~1NV, and masked/unmasked/\SI{25}{\percent}-cropped variants of Measurement~2NV.
    The synthetic curve shows a sharp corner characteristic of a well-posed inversion.
    The experimental curves broaden progressively, with Measurement~2NV displaced to significantly higher data-fidelity loss,
    consistent with its stronger ODMR sign-ambiguity artifacts.
    (b) Converged effective distance $d_{\mathrm{NV}}^\ast$ as a function of $\lambda$.
    For the synthetic case and Measurement~1NV, $d_{\mathrm{NV}}^\ast$ plateaus over a broad $\lambda$ range.
    For Measurement~2NV, the masked and unmasked variants drift monotonically,
    while the \SI{25}{\percent}-cropped variant recovers a plateau-like trace (Sec.~\ref{sec:model_data_consistency}).
    Filled squares mark the operating point $\lambda_\mathrm{bal}$ on each curve ($\lambda_\mathrm{bal}=10$ for the synthetic sweep, $\lambda_\mathrm{bal}=25$ for all experimental sweeps).}
    \label{fig:lcurve_comparison}
\end{figure*}

The L-curve can also serve as an internal consistency check between model and data. Since the synthetic dataset is inverted using its own forward model, its L-curve and $d_\mathrm{NV}^\ast(\lambda)$ trace represent the ideal, best-case reference. Both experimental datasets show a gap to this reference (Fig.~\ref{fig:lcurve_comparison}), which is consistent with real-world model deviations such as unmodeled $M_s$ inhomogeneities, film-thickness variations, or residual ODMR sign-ambiguity artefacts (Sec.~\ref{sec:odmr_artifacts}). It can, however, also contain contributions from measurement noise, the finite field of view, non-uniqueness, and incomplete optimization, and therefore flags a possible inconsistency without identifying its cause.

The two datasets also differ in behavior. Measurement~1NV is closer to the synthetic baseline, retaining a clear L-curve corner and a stable $d_\mathrm{NV}^\ast$ plateau (\num{80.2} to \SI{81.5}{\nano\metre} over $\lambda = 10$ to $160$). In contrast, Measurement~2NV exhibits a broader L-curve and a drifting $d_\mathrm{NV}^\ast$. This difference in behavior may stem from two key factors. First, Measurement~2NV is affected by more pronounced ODMR sign-ambiguity and measurement artefacts (Sec.~\ref{sec:odmr_artifacts}). Second, the smaller effective distance of Measurement~2NV probes the magnetization on sharper spatial scales, so any disagreement between our model and the actual sample is simply more visible.

To isolate the source of this instability, we analyze the masked, unmasked, and \SI{25}{\percent}-cropped variants of Measurement~2NV. The cropped variant excludes the bottom \SI{25}{\percent} of the scan area to remove both the large ODMR sign-ambiguity artifacts and the significant blurring in the bottom-right corner (whose physical origin is uncertain). As shown in Fig.~\ref{fig:lcurve_comparison}(b), masking already narrows the spread of the $d_\mathrm{NV}^\ast(\lambda)$ trace (\num{50} to \SI{76}{\nano\metre} masked against \num{42} to \SI{86}{\nano\metre} unmasked over the full sweep), and discarding the problematic region stabilizes the optimization further, yielding a plateau-like trace: over $\lambda = 10$ to $160$ the cropped variant spans \num{61.0} to \SI{63.8}{\nano\metre}, compared to \num{61.9} to \SI{66.2}{\nano\metre} masked and \num{61.4} to \SI{67.4}{\nano\metre} unmasked. This suggests that the dominant inconsistency is localized to specific non-ideal regions (e.g., boundary artifacts or local material defects) rather than representing a fundamental failure of the chosen physical model.

\subsection{Stability of Reconstructed Configurations}
\label{sec:llg_stability}
As an additional consistency check beyond field agreement, the stability of the optimized configurations $\mathbf{m}^\ast$ was evaluated under Landau--Lifshitz--Gilbert (LLG) relaxation~\cite{landau_theory_1935,gilbert_phenomenological_2004,bruckner_magnumnp_2023}. If these reconstructed states represented true equilibria, they would remain static, satisfying the condition that the magnetic torque vanishes:
\begin{equation}
  \mathbf{m} \times \mathbf{H}^{\mathrm{eff}} = 0,
\end{equation}
where the effective field $\mathbf{H}^{\mathrm{eff}}$ is the functional derivative of the total energy $E_{\mathrm{total}}$ with respect to the magnetization~\cite{abert_micromagnetics_2019}:
\begin{equation}
\mathbf{H}^{\mathrm{eff}} = -\frac{1}{\mu_0 M_s} \frac{\delta E_{\mathrm{total}}}{\delta \mathbf{m}}.
\end{equation}
The reconstructed structures do not abruptly collapse, indicating that they reside in a low-energy region of the state space, but they exhibit a slow spatial drift or gradual deformation as the system relaxes under the chosen energy functional. Some relaxation is expected by construction, since the reconstruction minimizes the joint loss of Eq.~\eqref{eq:loss_full} rather than the energy alone, so the optimized state need not be an exact energy minimum even for a perfectly specified model. Beyond this, the drift is consistent with minor discrepancies between the experimental sample (e.g., local pinning sites, thickness variations) and the simplified simulation model, which prevent the discovery of a strictly static solution under the assumed material parameters. The relaxation itself cannot distinguish such model error from spurious structure imprinted on $\mathbf{m}^\ast$ by measurement artifacts or noise. In either case, the reconstructed $\mathbf{m}^\ast$ is not a true micromagnetic equilibrium: for the assumed $\FGT$ material parameters we did not find a static configuration resembling the reconstructed skyrmion-lattice texture.

\section{Sensitivity and Robustness Study}
\label{sec:sensitivity_study}

To assess the reliability of the joint reconstruction, we examine how the converged effective distance $d_\mathrm{NV}^\ast$ responds to changes in the optimization starting point and in the assumed material parameters. All studies in this section use the experimental Measurement~1NV data.

\subsection{Robustness to the Initial Effective Distance}
\label{sec:sens_height_init}

We verify that the joint optimization of $d_\mathrm{NV}$ is insensitive to its initialization. Starting from four initial guesses $d_\mathrm{NV,0} \in \{40, 60, 80, 100\}\,\si{\nano\metre}$ at $\lambda = 25$, all runs converge to the same effective distance (Table~\ref{tab:height_init}).

\begin{table}[!ht]
\centering
\caption{\textbf{Robustness to the initial effective distance.} Converged effective distance $d_\mathrm{NV}^\ast$ at $\lambda = 25$ for four different initial guesses $d_\mathrm{NV,0}$.}
\label{tab:height_init}
\input{height_init_table.tex}
\end{table}

\subsection{Sensitivity to the Saturation Magnetization}
\label{sec:sens_ms}

\begin{figure}[!ht]
    \centering
    \resizebox{\columnwidth}{!}{\includegraphics{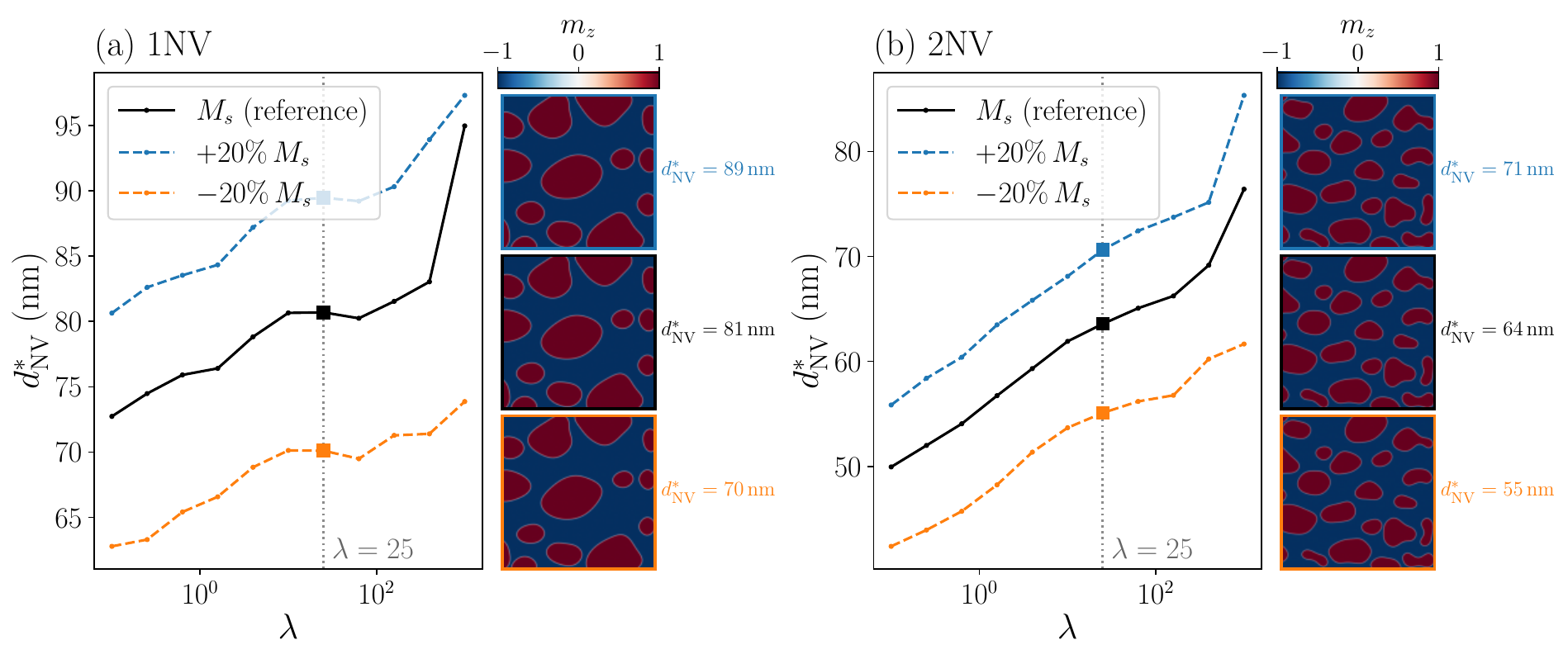}}
    \caption{\textbf{Sensitivity of the converged effective distance to the saturation magnetization.} (a)~1NV and (b)~2NV: converged effective distance $d_\mathrm{NV}^\ast$ versus $\lambda$ for the reference $M_s$ and $\pm\SI{20}{\percent}$ variants (squares mark $\lambda = 25$, dotted line), together with the reconstructed $m_z^\ast$ maps at that operating point. For 1NV, the reference $d_\mathrm{NV}^\ast = \SI{81}{\nano\metre}$ shifts to \SI{89}{\nano\metre} (\SI{+8}{\nano\metre}) for $+\SI{20}{\percent}\,M_s$ and to \SI{70}{\nano\metre} (\SI{-11}{\nano\metre}) for $-\SI{20}{\percent}\,M_s$; for 2NV the corresponding shifts are \SI{+7}{\nano\metre} and \SI{-9}{\nano\metre} from the reference of \SI{64}{\nano\metre}. Despite these $\sim\SI{10}{\percent}$--$\SI{14}{\percent}$ changes in the effective distance, the reconstructed domain texture is visually unchanged across all three $M_s$ values.}
    \label{fig:ms_sensitivity}
\end{figure}

Because the saturation magnetization $M_s$ directly determines the strength of the stray field compared with the measured values, it is probably the most influential model parameter for the effective distance optimization. Since $M_s$ is itself derived from a phenomenological domain-wall model (Eq.~\eqref{eq:Ms_derivation}), we quantify how an uncertainty in this value propagates into the reconstructed effective distance. We repeat the full L-curve sweep with $M_s$ varied by $\pm\SI{20}{\percent}$ relative to the reference value and compare the resulting $d_\mathrm{NV}^\ast(\lambda)$ curves in Fig.~\ref{fig:ms_sensitivity}. A $\pm\SI{20}{\percent}$ variation in $M_s$ shifts $d_\mathrm{NV}^\ast$ by \SI{10}{\percent}--\SI{14}{\percent} (\SI{8}{\nano\metre}--\SI{11}{\nano\metre} for 1NV, \SI{7}{\nano\metre}--\SI{9}{\nano\metre} for 2NV), but at the same $\lambda$ value all three $M_s$ variants recover the same domain texture.

\subsection{Sensitivity to the Film Thickness}
\label{sec:sens_thickness}

\begin{figure}[!ht]
    \centering
    \resizebox{\columnwidth}{!}{\includegraphics{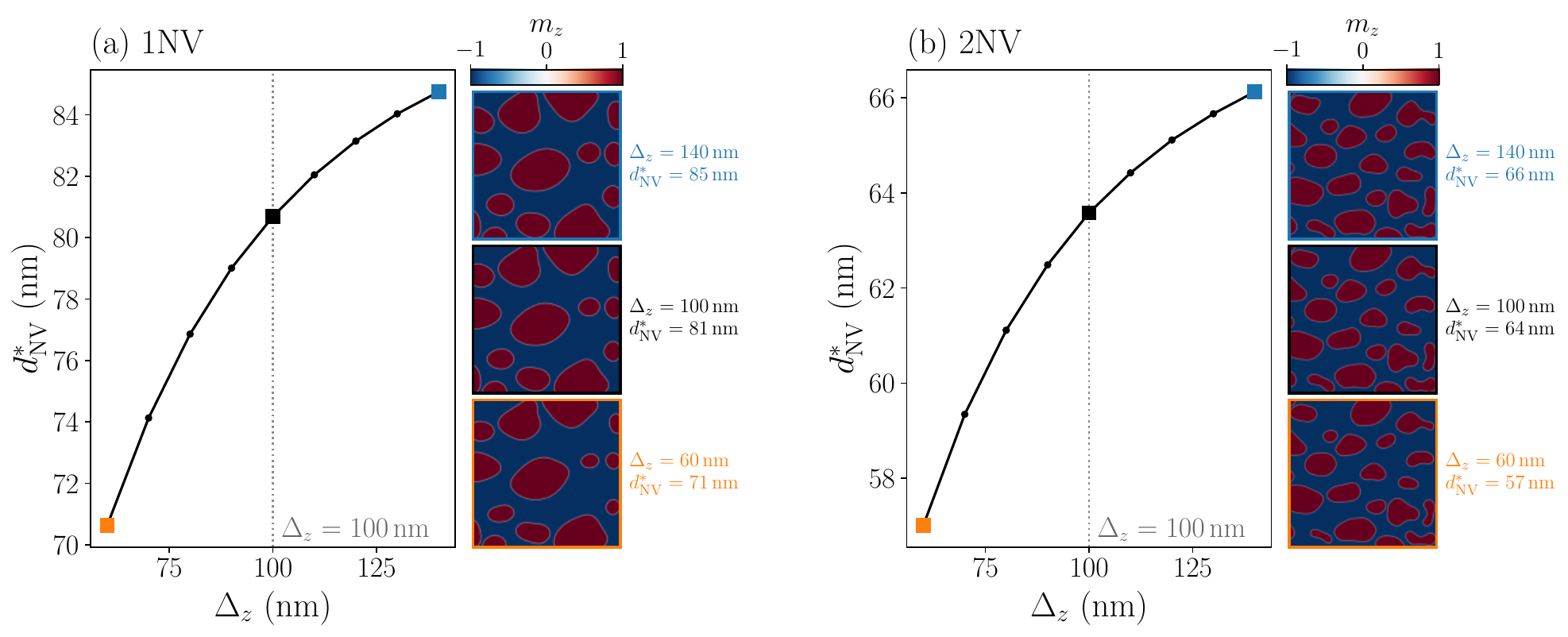}}
    \caption{\textbf{Sensitivity of the converged effective distance to the film thickness.} (a)~1NV and (b)~2NV: converged effective distance $d_\mathrm{NV}^\ast$ as a function of the layer thickness $\Delta_z$ at $\lambda = 25$, with the reconstructed $m_z^\ast$ maps at $\Delta_z = \SI{140}{}$, \SI{100}{} (nominal, dotted line) and \SI{60}{\nano\metre} (squares). For 1NV the nominal $d_\mathrm{NV}^\ast = \SI{81}{\nano\metre}$ ranges from \SI{71}{\nano\metre} at \SI{60}{\nano\metre} to \SI{85}{\nano\metre} at \SI{140}{\nano\metre}; for 2NV the reference \SI{64}{\nano\metre} ranges from \SI{57}{} to \SI{66}{\nano\metre}. The dependence is strongly nonlinear and saturates above the nominal thickness, and the texture is preserved across the range, with the domains slightly smaller at the thin end.}
    \label{fig:thickness_sensitivity}
\end{figure}

The magnetic layer is modelled as a single cell of thickness $\Delta_z = \SI{100}{\nano\metre}$, equal to the nominal flake thickness. Since the stray field is set by the product $M_s\,\Delta_z$, an uncertainty in the thickness affects the reconstructed effective distance similarly to an uncertainty in $M_s$. To quantify this, we repeat the reconstruction at the operating point $\lambda = 25$ for thicknesses $\Delta_z$ between \SI{60}{} and \SI{140}{\nano\metre} and record the converged $d_\mathrm{NV}^\ast$ (Fig.~\ref{fig:thickness_sensitivity}).

The converged $d_\mathrm{NV}^\ast$ increases monotonically with $\Delta_z$, but the dependence is strongly nonlinear. Near the nominal thickness it is weak: a $\pm\SI{20}{\percent}$ change in $\Delta_z$ shifts $d_\mathrm{NV}^\ast$ by less than \SI{5}{\percent} for both datasets, about a third of the shift caused by the same relative change in $M_s$. The change is much larger at the thin end, where a step from \SI{60}{} to \SI{70}{\nano\metre} moves $d_\mathrm{NV}^\ast$ far more than a step from \SI{110}{} to \SI{120}{\nano\metre}. This is intuitive, since for a thin film the material sits closer to the sensor, so an added layer has a stronger effect on the measured field than the same layer added to a thicker film further away. The domain configuration is preserved over the whole range, with the reconstructed domains becoming slightly smaller toward the thin end.

\subsection{Sensitivity to the Magnetization Initialization}
\label{sec:sens_init}

\begin{figure*}[!ht]
    \centering
    \resizebox{\textwidth}{!}{\includegraphics{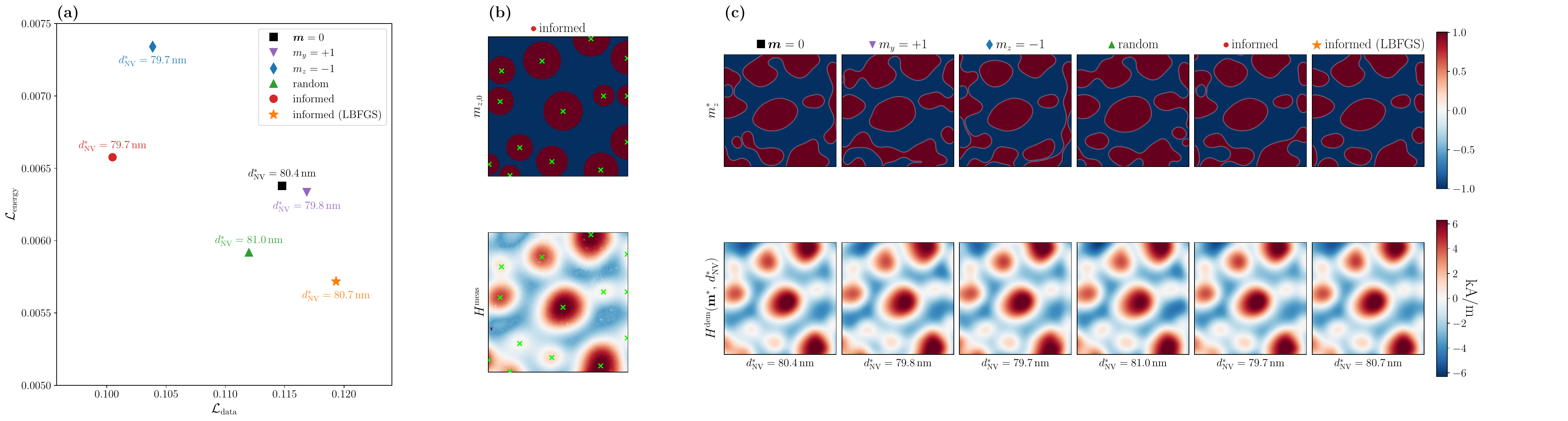}}
    \caption{\textbf{Sensitivity to the magnetization initialization.}
    Reconstructions are performed with the Riemannian Adam (R-Adam) optimizer starting from five different initializations: zero ($\mathbf{m} = 0$), uniform out-of-plane ($m_z = -1$), uniform in-plane ($m_y = +1$), random, or informed (derived from the $H^{\mathrm{meas}}$ peaks shown in (b)). For the informed initialization, the L-BFGS reconstruction from the main text is added for comparison.
    (a) shows each final reconstruction in the phase plot ($\mathcal{L}_\mathrm{energy}$ versus $\mathcal{L}_\mathrm{data}$ at $\lambda = 10$).
    (b) shows the detected peaks in the measured field $H^{\mathrm{meas}}$ and the resulting initial out-of-plane magnetization $m_{z,0}$ used for the informed case.
    (c) shows the difference in the reconstructed out-of-plane magnetization $m^\ast_z$ structure (top row) despite very similar simulated stray fields $H^{\mathrm{dem}}(\mathbf{m}^\ast,d^\ast_\mathrm{NV})$ (bottom row).}
    \label{fig:init_sensitivity}
\end{figure*}

Finally, we examine the dependence of the reconstruction on the initial magnetization state. Using the same Riemannian-Adam optimizer throughout, we compare five starting configurations: an informed initialization in which positive out-of-plane domains are placed at peaks detected in the measured field $H^{\mathrm{meas}}$ (Fig.~\ref{fig:init_sensitivity}(b)), a cold start ($\mathbf{m} = \mathbf{0}$), a random configuration, a uniform in-plane state ($m_y = +1$), and a uniform out-of-plane state ($m_z = -1$). The informed initialization is additionally run with the second-order L-BFGS optimizer of the main text, giving six runs in total.

All six runs converge to a consistent effective distance and reproduce the measured field to a comparable degree (Fig.~\ref{fig:init_sensitivity}(a)). The reconstructed $m^\ast_z$ textures, however, differ visibly at small scales (Fig.~\ref{fig:init_sensitivity}(c)), a direct consequence of the non-uniqueness of the inverse problem, which persists even after the micromagnetic energy is added to the loss. The large-scale reconstruction is therefore robust across the initializations compared here, while the fine-scale structure remains underdetermined in this case. Because the precise outcome of each run depends on the numerical setup (optimizer, initialization, and regularization strength) rather than being a universal property of the framework, we rank the candidates with the $(\mathcal{L}_\mathrm{data}, \mathcal{L}_\mathrm{energy})$ phase plot, preferring the lower-energy configuration at equal data fit as in the main text. The informed L-BFGS reconstruction of the main text is adopted on this basis.

\section{Physical Composition of the Extracted Effective Distances}
\label{sec:dnv_composition}

\begin{figure}[t]
    \centering
    \resizebox{0.62\textwidth}{!}{\includegraphics{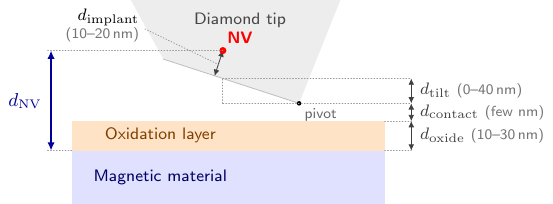}}
    \caption{\textbf{Physical composition of the effective distance $d_\mathrm{NV}$.} Schematic (not to scale) of the contributions to the NV--material separation: the implantation depth $d_\mathrm{implant}$ below the diamond side, the mechanical contact gap $d_\mathrm{contact}$, the tilt offset $d_\mathrm{tilt}$ that lifts the laterally offset NV above the contact pivot, and the sample oxidation layer $d_\mathrm{oxide}$. The implantation depth is measured normal to the locally tilted side, so the listed ranges sum to the vertical $d_\mathrm{NV}$ only up to a $\cos$ factor of the small tilt angle.}
    \label{fig:dnv_sketch}
\end{figure}

The effective distance is the total separation between the magnetic material and the NV sensing center,
\begin{equation}
d_\mathrm{NV} = d_\mathrm{implant} + d_\mathrm{contact} + d_\mathrm{oxide} + d_\mathrm{tilt},
\label{eq:dnv_decomposition}
\end{equation}
comprising the NV implantation depth in the diamond, the surface oxidation layer of the sample, and the physical sample--diamond separation, which we split into the mechanical contact gap and the tilt-induced geometric offset (Fig.~\ref{fig:dnv_sketch}). These contributions are not directly measured for the present setup. We therefore estimate them from the fabrication and setup parameters, guided by the systematic characterization of comparable scanning-NV probes by Xu~\textit{et~al.}~\cite{xu_minimizing_2025}.

\emph{Implantation depth} ($d_\mathrm{implant}$). The NV centers are implanted at a beam energy of \SI{6}{\kilo\electronvolt}, which we estimate places them \num{10} to \SI{20}{\nano\metre} below the locally tilted diamond side. Xu~\textit{et~al.} measure NV depths of \num{5} to \SI{14}{\nano\metre} (median \SI{9}{\nano\metre}) for similar tips implanted at \SI{7}{\kilo\electronvolt}~\cite{xu_minimizing_2025}.

\emph{Contact gap} ($d_\mathrm{contact}$). In AFM contact the mechanical gap at the contact point is only a few nanometres in the ideal case, but surface features on the diamond tip, such as lithography residues with peak heights of \num{20} to \SI{50}{\nano\metre} or material picked up during scanning, as well as adsorbate layers, can prevent a closer approach and enlarge the effective distance~\cite{xu_minimizing_2025}.

\emph{Tilt offset} ($d_\mathrm{tilt}$). The bottom of the diamond tip is \num{200} to \SI{400}{\nano\metre} wide and the NV can lie anywhere across it. A relative tilt between tip and sample makes one edge the contact pivot, so a laterally offset NV is held above the contact point, which we estimate adds \num{0} to \SI{40}{\nano\metre}. The same part of the tip stays in contact throughout a scan, so this offset is constant within a measurement.

\emph{Oxidation layer} ($d_\mathrm{oxide}$). The uncapped, air-sensitive flakes develop a surface oxide that acts as a magnetically dead spacer. On the closely related $\mathrm{Fe_3GeTe_2}$, this air-formed oxide thickens progressively over weeks of ambient exposure~\cite{xie_air_2024}. For the present flakes, which show progressive degradation over months of storage, we estimate \num{10} to \SI{30}{\nano\metre}.

Taken together, these estimates are consistent with the \num{26} to \SI{128}{\nano\metre} effective NV--sample distances (medians \num{43} and \SI{60}{\nano\metre} for the two feedback modes) measured across 15 scanning-NV probes by Xu~\textit{et~al.}~\cite{xu_minimizing_2025}. Both extracted values, $d_{\mathrm{NV}}^\ast \approx \SI{81}{\nano\metre}$ for Measurement~1NV and $d_{\mathrm{NV}}^\ast \approx \SI{64}{\nano\metre}$ for Measurement~2NV, fall within this range.

Both scans used the same tip and were taken about a week apart, on different $\FGT$ flakes from the same exfoliation on a common substrate. Since the tip was not exchanged, the implantation depth is fixed and a tip change can be ruled out as the origin of the \SI{17}{\nano\metre} difference. The remaining contributions are sample- and contact-dependent. The two flakes can differ in thickness, and since the tip was retracted and re-approached for each scan, the contact geometry and the associated tilt offset may not be identical between the two sessions. The later scan (Measurement~1NV) also carries the larger extracted distance, which matches the direction expected for progressive surface oxidation under ambient air storage, although a \SI{17}{\nano\metre} change within a week exceeds what is documented for the related $\mathrm{Fe_3GeTe_2}$, and no oxidation data exist for $\FGT$ itself. We therefore attribute the difference to a combination of these sample- and contact-dependent effects.

\end{document}

%% file: height_init_table.tex
\begin{tabular}{cc}
\hline\hline
$d_\mathrm{NV,0}$ (nm) & $d_\mathrm{NV}^\ast$ (nm) \\
\hline
40 & 80.70 \\
60 & 80.70 \\
80 & 80.69 \\
100 & 80.70 \\
\hline\hline
\end{tabular}

%% file: main.bbl
\begin{thebibliography}{34}%
\makeatletter
\providecommand \@ifxundefined [1]{%
 \@ifx{#1\undefined}
}%
\providecommand \@ifnum [1]{%
 \ifnum #1\expandafter \@firstoftwo
 \else \expandafter \@secondoftwo
 \fi
}%
\providecommand \@ifx [1]{%
 \ifx #1\expandafter \@firstoftwo
 \else \expandafter \@secondoftwo
 \fi
}%
\providecommand \natexlab [1]{#1}%
\providecommand \enquote  [1]{``#1''}%
\providecommand \bibnamefont  [1]{#1}%
\providecommand \bibfnamefont [1]{#1}%
\providecommand \citenamefont [1]{#1}%
\providecommand \href@noop [0]{\@secondoftwo}%
\providecommand \href [0]{\begingroup \@sanitize@url \@href}%
\providecommand \@href[1]{\@@startlink{#1}\@@href}%
\providecommand \@@href[1]{\endgroup#1\@@endlink}%
\providecommand \@sanitize@url [0]{\catcode `\\12\catcode `\$12\catcode
  `\&12\catcode `\#12\catcode `\^12\catcode `\_12\catcode `\%12\relax}%
\providecommand \@@startlink[1]{}%
\providecommand \@@endlink[0]{}%
\providecommand \url  [0]{\begingroup\@sanitize@url \@url }%
\providecommand \@url [1]{\endgroup\@href {#1}{\urlprefix }}%
\providecommand \urlprefix  [0]{URL }%
\providecommand \Eprint [0]{\href }%
\providecommand \doibase [0]{https://doi.org/}%
\providecommand \selectlanguage [0]{\@gobble}%
\providecommand \bibinfo  [0]{\@secondoftwo}%
\providecommand \bibfield  [0]{\@secondoftwo}%
\providecommand \translation [1]{[#1]}%
\providecommand \BibitemOpen [0]{}%
\providecommand \bibitemStop [0]{}%
\providecommand \bibitemNoStop [0]{.\EOS\space}%
\providecommand \EOS [0]{\spacefactor3000\relax}%
\providecommand \BibitemShut  [1]{\csname bibitem#1\endcsname}%
\let\auto@bib@innerbib\@empty
\bibitem [{\citenamefont {Casola}\ \emph {et~al.}(2018)\citenamefont {Casola},
  \citenamefont {van~der Sar},\ and\ \citenamefont
  {Yacoby}}]{casola_probing_2018}%
  \BibitemOpen
  \bibfield  {author} {\bibinfo {author} {\bibfnamefont {F.}~\bibnamefont
  {Casola}}, \bibinfo {author} {\bibfnamefont {T.}~\bibnamefont {van~der
  Sar}},\ and\ \bibinfo {author} {\bibfnamefont {A.}~\bibnamefont {Yacoby}},\
  }\bibfield  {title} {\bibinfo {title} {Probing condensed matter physics with
  magnetometry based on nitrogen-vacancy centres in diamond},\ }\href
  {https://doi.org/10.1038/natrevmats.2017.88} {\bibfield  {journal} {\bibinfo
  {journal} {Nature Reviews Materials}\ }\textbf {\bibinfo {volume} {3}},\
  \bibinfo {pages} {17088} (\bibinfo {year} {2018})}\BibitemShut {NoStop}%
\bibitem [{\citenamefont {Xu}\ \emph {et~al.}(2023)\citenamefont {Xu},
  \citenamefont {Zhang},\ and\ \citenamefont {Tian}}]{xu_recent_2023}%
  \BibitemOpen
  \bibfield  {author} {\bibinfo {author} {\bibfnamefont {Y.}~\bibnamefont
  {Xu}}, \bibinfo {author} {\bibfnamefont {W.}~\bibnamefont {Zhang}},\ and\
  \bibinfo {author} {\bibfnamefont {C.}~\bibnamefont {Tian}},\ }\bibfield
  {title} {\bibinfo {title} {Recent advances on applications of
  $\text{NV}^{\textrm{-}}$ magnetometry in condensed matter physics},\ }\href
  {https://doi.org/10.1364/PRJ.471266} {\bibfield  {journal} {\bibinfo
  {journal} {Photonics Research}\ }\textbf {\bibinfo {volume} {11}},\ \bibinfo
  {pages} {393} (\bibinfo {year} {2023})}\BibitemShut {NoStop}%
\bibitem [{\citenamefont {Backus}(1970)}]{backus_non-uniqueness_1970}%
  \BibitemOpen
  \bibfield  {author} {\bibinfo {author} {\bibfnamefont {G.~E.}\ \bibnamefont
  {Backus}},\ }\bibfield  {title} {\bibinfo {title} {Non-uniqueness of the
  external geomagnetic field determined by surface intensity measurements},\
  }\href {https://doi.org/10.1029/JA075i031p06339} {\bibfield  {journal}
  {\bibinfo  {journal} {Journal of Geophysical Research (1896-1977)}\ }\textbf
  {\bibinfo {volume} {75}},\ \bibinfo {pages} {6339} (\bibinfo {year}
  {1970})},\ \bibinfo {note} {\_eprint:
  https://agupubs.onlinelibrary.wiley.com/doi/pdf/10.1029/JA075i031p06339}\BibitemShut
  {NoStop}%
\bibitem [{\citenamefont {Bruckner}\ \emph {et~al.}(2017)\citenamefont
  {Bruckner}, \citenamefont {Abert}, \citenamefont {Wautischer}, \citenamefont
  {Huber}, \citenamefont {Vogler}, \citenamefont {Hinze},\ and\ \citenamefont
  {Suess}}]{bruckner_solving_2017}%
  \BibitemOpen
  \bibfield  {author} {\bibinfo {author} {\bibfnamefont {F.}~\bibnamefont
  {Bruckner}}, \bibinfo {author} {\bibfnamefont {C.}~\bibnamefont {Abert}},
  \bibinfo {author} {\bibfnamefont {G.}~\bibnamefont {Wautischer}}, \bibinfo
  {author} {\bibfnamefont {C.}~\bibnamefont {Huber}}, \bibinfo {author}
  {\bibfnamefont {C.}~\bibnamefont {Vogler}}, \bibinfo {author} {\bibfnamefont
  {M.}~\bibnamefont {Hinze}},\ and\ \bibinfo {author} {\bibfnamefont
  {D.}~\bibnamefont {Suess}},\ }\bibfield  {title} {\bibinfo {title} {Solving
  {Large}-{Scale} {Inverse} {Magnetostatic} {Problems} using the {Adjoint}
  {Method}},\ }\href {https://doi.org/10.1038/srep40816} {\bibfield  {journal}
  {\bibinfo  {journal} {Scientific Reports}\ }\textbf {\bibinfo {volume} {7}},\
  \bibinfo {pages} {40816} (\bibinfo {year} {2017})}\BibitemShut {NoStop}%
\bibitem [{\citenamefont {Clement}\ \emph {et~al.}(2021)\citenamefont
  {Clement}, \citenamefont {Sethna},\ and\ \citenamefont
  {Nowack}}]{clement_reconstruction_2021}%
  \BibitemOpen
  \bibfield  {author} {\bibinfo {author} {\bibfnamefont {C.~B.}\ \bibnamefont
  {Clement}}, \bibinfo {author} {\bibfnamefont {J.~P.}\ \bibnamefont
  {Sethna}},\ and\ \bibinfo {author} {\bibfnamefont {K.~C.}\ \bibnamefont
  {Nowack}},\ }\href {https://doi.org/10.48550/arXiv.1910.12929} {\bibinfo
  {title} {Reconstruction of {Current} {Densities} from {Magnetic} {Images} by
  {Bayesian} {Inference}}} (\bibinfo {year} {2021}),\ \bibinfo {note}
  {arXiv:1910.12929 [physics]}\BibitemShut {NoStop}%
\bibitem [{\citenamefont {Hug}\ \emph {et~al.}(1998)\citenamefont {Hug},
  \citenamefont {Stiefel}, \citenamefont {van Schendel}, \citenamefont {Moser},
  \citenamefont {Hofer}, \citenamefont {Martin}, \citenamefont {Güntherodt},
  \citenamefont {Porthun}, \citenamefont {Abelmann}, \citenamefont {Lodder},
  \citenamefont {Bochi},\ and\ \citenamefont
  {O’Handley}}]{hug_quantitative_1998}%
  \BibitemOpen
  \bibfield  {author} {\bibinfo {author} {\bibfnamefont {H.~J.}\ \bibnamefont
  {Hug}}, \bibinfo {author} {\bibfnamefont {B.}~\bibnamefont {Stiefel}},
  \bibinfo {author} {\bibfnamefont {P.~J.~A.}\ \bibnamefont {van Schendel}},
  \bibinfo {author} {\bibfnamefont {A.}~\bibnamefont {Moser}}, \bibinfo
  {author} {\bibfnamefont {R.}~\bibnamefont {Hofer}}, \bibinfo {author}
  {\bibfnamefont {S.}~\bibnamefont {Martin}}, \bibinfo {author} {\bibfnamefont
  {H.-J.}\ \bibnamefont {Güntherodt}}, \bibinfo {author} {\bibfnamefont
  {S.}~\bibnamefont {Porthun}}, \bibinfo {author} {\bibfnamefont
  {L.}~\bibnamefont {Abelmann}}, \bibinfo {author} {\bibfnamefont {J.~C.}\
  \bibnamefont {Lodder}}, \bibinfo {author} {\bibfnamefont {G.}~\bibnamefont
  {Bochi}},\ and\ \bibinfo {author} {\bibfnamefont {R.~C.}\ \bibnamefont
  {O’Handley}},\ }\bibfield  {title} {\bibinfo {title} {Quantitative magnetic
  force microscopy on perpendicularly magnetized samples},\ }\href
  {https://doi.org/10.1063/1.367412} {\bibfield  {journal} {\bibinfo  {journal}
  {Journal of Applied Physics}\ }\textbf {\bibinfo {volume} {83}},\ \bibinfo
  {pages} {5609} (\bibinfo {year} {1998})}\BibitemShut {NoStop}%
\bibitem [{\citenamefont {Feng}\ \emph {et~al.}(2022)\citenamefont {Feng},
  \citenamefont {Mandru}, \citenamefont {Yıldırım},\ and\ \citenamefont
  {Hug}}]{feng_quantitative_2022}%
  \BibitemOpen
  \bibfield  {author} {\bibinfo {author} {\bibfnamefont {Y.}~\bibnamefont
  {Feng}}, \bibinfo {author} {\bibfnamefont {A.-O.}\ \bibnamefont {Mandru}},
  \bibinfo {author} {\bibfnamefont {O.}~\bibnamefont {Yıldırım}},\ and\
  \bibinfo {author} {\bibfnamefont {H.}~\bibnamefont {Hug}},\ }\bibfield
  {title} {\bibinfo {title} {Quantitative {Magnetic} {Force} {Microscopy}:
  {Transfer}-{Function} {Method} {Revisited}},\ }\href
  {https://doi.org/10.1103/PhysRevApplied.18.024016} {\bibfield  {journal}
  {\bibinfo  {journal} {Physical Review Applied}\ }\textbf {\bibinfo {volume}
  {18}},\ \bibinfo {pages} {024016} (\bibinfo {year} {2022})}\BibitemShut
  {NoStop}%
\bibitem [{\citenamefont {Yao}\ \emph {et~al.}(2025)\citenamefont {Yao},
  \citenamefont {Yu}, \citenamefont {Shi}, \citenamefont {Jung}, \citenamefont
  {Váci}, \citenamefont {Wang}, \citenamefont {Liu}, \citenamefont {Zhang},
  \citenamefont {Tikoo-Schantz},\ and\ \citenamefont
  {Zu}}]{yao_universal_2025}%
  \BibitemOpen
  \bibfield  {author} {\bibinfo {author} {\bibfnamefont {C.}~\bibnamefont
  {Yao}}, \bibinfo {author} {\bibfnamefont {Y.}~\bibnamefont {Yu}}, \bibinfo
  {author} {\bibfnamefont {Y.}~\bibnamefont {Shi}}, \bibinfo {author}
  {\bibfnamefont {J.-I.}\ \bibnamefont {Jung}}, \bibinfo {author}
  {\bibfnamefont {Z.}~\bibnamefont {Váci}}, \bibinfo {author} {\bibfnamefont
  {Y.}~\bibnamefont {Wang}}, \bibinfo {author} {\bibfnamefont {Z.}~\bibnamefont
  {Liu}}, \bibinfo {author} {\bibfnamefont {C.}~\bibnamefont {Zhang}}, \bibinfo
  {author} {\bibfnamefont {S.}~\bibnamefont {Tikoo-Schantz}},\ and\ \bibinfo
  {author} {\bibfnamefont {C.}~\bibnamefont {Zu}},\ }\bibfield  {title}
  {\bibinfo {title} {Universal reconstruction of complex magnetic profiles with
  minimal prior assumptions},\ }\href {https://doi.org/10.1103/q312-kf83}
  {\bibfield  {journal} {\bibinfo  {journal} {Physical Review Applied}\
  }\textbf {\bibinfo {volume} {24}},\ \bibinfo {pages} {064020} (\bibinfo
  {year} {2025})}\BibitemShut {NoStop}%
\bibitem [{\citenamefont {Broadway}\ \emph {et~al.}(2025)\citenamefont
  {Broadway}, \citenamefont {Flaks}, \citenamefont {Dubois},\ and\
  \citenamefont {Maletinsky}}]{broadway_reconstruction_2025}%
  \BibitemOpen
  \bibfield  {author} {\bibinfo {author} {\bibfnamefont {D.~A.}\ \bibnamefont
  {Broadway}}, \bibinfo {author} {\bibfnamefont {M.}~\bibnamefont {Flaks}},
  \bibinfo {author} {\bibfnamefont {A.~E.}\ \bibnamefont {Dubois}},\ and\
  \bibinfo {author} {\bibfnamefont {P.}~\bibnamefont {Maletinsky}},\ }\bibfield
   {title} {\bibinfo {title} {Reconstruction of nontrivial magnetization
  textures from magnetic field images using neural networks},\ }\href
  {https://doi.org/10.1103/PhysRevApplied.23.044012} {\bibfield  {journal}
  {\bibinfo  {journal} {Physical Review Applied}\ }\textbf {\bibinfo {volume}
  {23}},\ \bibinfo {pages} {044012} (\bibinfo {year} {2025})}\BibitemShut
  {NoStop}%
\bibitem [{\citenamefont {Dubois}\ \emph {et~al.}(2022)\citenamefont {Dubois},
  \citenamefont {Broadway}, \citenamefont {Stark}, \citenamefont {Tschudin},
  \citenamefont {Healey}, \citenamefont {Huber}, \citenamefont {Tetienne},
  \citenamefont {Greplova},\ and\ \citenamefont
  {Maletinsky}}]{dubois_untrained_2022}%
  \BibitemOpen
  \bibfield  {author} {\bibinfo {author} {\bibfnamefont {A.}~\bibnamefont
  {Dubois}}, \bibinfo {author} {\bibfnamefont {D.}~\bibnamefont {Broadway}},
  \bibinfo {author} {\bibfnamefont {A.}~\bibnamefont {Stark}}, \bibinfo
  {author} {\bibfnamefont {M.}~\bibnamefont {Tschudin}}, \bibinfo {author}
  {\bibfnamefont {A.}~\bibnamefont {Healey}}, \bibinfo {author} {\bibfnamefont
  {S.}~\bibnamefont {Huber}}, \bibinfo {author} {\bibfnamefont {J.-P.}\
  \bibnamefont {Tetienne}}, \bibinfo {author} {\bibfnamefont {E.}~\bibnamefont
  {Greplova}},\ and\ \bibinfo {author} {\bibfnamefont {P.}~\bibnamefont
  {Maletinsky}},\ }\bibfield  {title} {\bibinfo {title} {Untrained {Physically}
  {Informed} {Neural} {Network} for {Image} {Reconstruction} of {Magnetic}
  {Field} {Sources}},\ }\href
  {https://doi.org/10.1103/PhysRevApplied.18.064076} {\bibfield  {journal}
  {\bibinfo  {journal} {Physical Review Applied}\ }\textbf {\bibinfo {volume}
  {18}},\ \bibinfo {pages} {064076} (\bibinfo {year} {2022})}\BibitemShut
  {NoStop}%
\bibitem [{\citenamefont {Suess}\ \emph {et~al.}(2025)\citenamefont {Suess},
  \citenamefont {Setescak}, \citenamefont {Smith}, \citenamefont {Bruckner},
  \citenamefont {Hug},\ and\ \citenamefont
  {Abert}}]{suess_reconstruction_2025}%
  \BibitemOpen
  \bibfield  {author} {\bibinfo {author} {\bibfnamefont {D.}~\bibnamefont
  {Suess}}, \bibinfo {author} {\bibfnamefont {A.}~\bibnamefont {Setescak}},
  \bibinfo {author} {\bibfnamefont {J.}~\bibnamefont {Smith}}, \bibinfo
  {author} {\bibfnamefont {F.}~\bibnamefont {Bruckner}}, \bibinfo {author}
  {\bibfnamefont {H.~J.}\ \bibnamefont {Hug}},\ and\ \bibinfo {author}
  {\bibfnamefont {C.}~\bibnamefont {Abert}},\ }\bibfield  {title} {\bibinfo
  {title} {Reconstruction of magnetic structures and material parameters with
  convolutional neural network and bias field-constrained micromagnetic
  relaxation},\ }\href {https://doi.org/10.1038/s41598-025-27151-1} {\bibfield
  {journal} {\bibinfo  {journal} {Scientific Reports}\ }\textbf {\bibinfo
  {volume} {15}},\ \bibinfo {pages} {42867} (\bibinfo {year}
  {2025})}\BibitemShut {NoStop}%
\bibitem [{\citenamefont {Hansen}\ and\ \citenamefont
  {O’Leary}(1993)}]{hansen_use_1993}%
  \BibitemOpen
  \bibfield  {author} {\bibinfo {author} {\bibfnamefont {P.~C.}\ \bibnamefont
  {Hansen}}\ and\ \bibinfo {author} {\bibfnamefont {D.~P.}\ \bibnamefont
  {O’Leary}},\ }\bibfield  {title} {\bibinfo {title} {The {Use} of the
  {L}-{Curve} in the {Regularization} of {Discrete} {Ill}-{Posed} {Problems}},\
  }\href {https://doi.org/10.1137/0914086} {\bibfield  {journal} {\bibinfo
  {journal} {SIAM Journal on Scientific Computing}\ }\textbf {\bibinfo {volume}
  {14}},\ \bibinfo {pages} {1487} (\bibinfo {year} {1993})}\BibitemShut
  {NoStop}%
\bibitem [{\citenamefont {Calvetti}\ \emph {et~al.}(2000)\citenamefont
  {Calvetti}, \citenamefont {Morigi}, \citenamefont {Reichel},\ and\
  \citenamefont {Sgallari}}]{calvetti_tikhonov_2000}%
  \BibitemOpen
  \bibfield  {author} {\bibinfo {author} {\bibfnamefont {D.}~\bibnamefont
  {Calvetti}}, \bibinfo {author} {\bibfnamefont {S.}~\bibnamefont {Morigi}},
  \bibinfo {author} {\bibfnamefont {L.}~\bibnamefont {Reichel}},\ and\ \bibinfo
  {author} {\bibfnamefont {F.}~\bibnamefont {Sgallari}},\ }\bibfield  {title}
  {\bibinfo {title} {Tikhonov regularization and the {L}-curve for large
  discrete ill-posed problems},\ }\href
  {https://doi.org/10.1016/S0377-0427(00)00414-3} {\bibfield  {journal}
  {\bibinfo  {journal} {Journal of Computational and Applied Mathematics}\
  }\bibinfo {series} {Numerical {Analysis} 2000. {Vol}. {III}: {Linear}
  {Algebra}},\ \textbf {\bibinfo {volume} {123}},\ \bibinfo {pages} {423}
  (\bibinfo {year} {2000})}\BibitemShut {NoStop}%
\bibitem [{\citenamefont {Xu}\ \emph {et~al.}(2025)\citenamefont {Xu},
  \citenamefont {Palm}, \citenamefont {Huxter}, \citenamefont {Herb},
  \citenamefont {Abendroth}, \citenamefont {Bouzehouane}, \citenamefont
  {Boulle}, \citenamefont {Gabor}, \citenamefont {Urrestarazu~Larranaga},
  \citenamefont {Morales}, \citenamefont {Rhensius}, \citenamefont
  {Puebla-Hellmann},\ and\ \citenamefont {Degen}}]{xu_minimizing_2025}%
  \BibitemOpen
  \bibfield  {author} {\bibinfo {author} {\bibfnamefont {Z.}~\bibnamefont
  {Xu}}, \bibinfo {author} {\bibfnamefont {M.~L.}\ \bibnamefont {Palm}},
  \bibinfo {author} {\bibfnamefont {W.}~\bibnamefont {Huxter}}, \bibinfo
  {author} {\bibfnamefont {K.}~\bibnamefont {Herb}}, \bibinfo {author}
  {\bibfnamefont {J.~M.}\ \bibnamefont {Abendroth}}, \bibinfo {author}
  {\bibfnamefont {K.}~\bibnamefont {Bouzehouane}}, \bibinfo {author}
  {\bibfnamefont {O.}~\bibnamefont {Boulle}}, \bibinfo {author} {\bibfnamefont
  {M.~S.}\ \bibnamefont {Gabor}}, \bibinfo {author} {\bibfnamefont
  {J.}~\bibnamefont {Urrestarazu~Larranaga}}, \bibinfo {author} {\bibfnamefont
  {A.}~\bibnamefont {Morales}}, \bibinfo {author} {\bibfnamefont
  {J.}~\bibnamefont {Rhensius}}, \bibinfo {author} {\bibfnamefont
  {G.}~\bibnamefont {Puebla-Hellmann}},\ and\ \bibinfo {author} {\bibfnamefont
  {C.~L.}\ \bibnamefont {Degen}},\ }\bibfield  {title} {\bibinfo {title}
  {Minimizing {Sensor}-{Sample} {Distances} in {Scanning} {Nitrogen}-{Vacancy}
  {Magnetometry}},\ }\href {https://doi.org/10.1021/acsnano.4c18460} {\bibfield
   {journal} {\bibinfo  {journal} {ACS Nano}\ }\textbf {\bibinfo {volume}
  {19}},\ \bibinfo {pages} {8255} (\bibinfo {year} {2025})}\BibitemShut
  {NoStop}%
\bibitem [{\citenamefont {Abert}\ \emph {et~al.}(2025)\citenamefont {Abert},
  \citenamefont {Bruckner}, \citenamefont {Voronov}, \citenamefont {Lang},
  \citenamefont {Pathak}, \citenamefont {Holt}, \citenamefont {Kraft},
  \citenamefont {Allayarov}, \citenamefont {Flauger}, \citenamefont {Koraltan},
  \citenamefont {Schrefl}, \citenamefont {Chumak}, \citenamefont {Fangohr},\
  and\ \citenamefont {Suess}}]{abert_neuralmag_2025}%
  \BibitemOpen
  \bibfield  {author} {\bibinfo {author} {\bibfnamefont {C.}~\bibnamefont
  {Abert}}, \bibinfo {author} {\bibfnamefont {F.}~\bibnamefont {Bruckner}},
  \bibinfo {author} {\bibfnamefont {A.}~\bibnamefont {Voronov}}, \bibinfo
  {author} {\bibfnamefont {M.}~\bibnamefont {Lang}}, \bibinfo {author}
  {\bibfnamefont {S.~A.}\ \bibnamefont {Pathak}}, \bibinfo {author}
  {\bibfnamefont {S.}~\bibnamefont {Holt}}, \bibinfo {author} {\bibfnamefont
  {R.}~\bibnamefont {Kraft}}, \bibinfo {author} {\bibfnamefont
  {R.}~\bibnamefont {Allayarov}}, \bibinfo {author} {\bibfnamefont
  {P.}~\bibnamefont {Flauger}}, \bibinfo {author} {\bibfnamefont
  {S.}~\bibnamefont {Koraltan}}, \bibinfo {author} {\bibfnamefont
  {T.}~\bibnamefont {Schrefl}}, \bibinfo {author} {\bibfnamefont
  {A.}~\bibnamefont {Chumak}}, \bibinfo {author} {\bibfnamefont
  {H.}~\bibnamefont {Fangohr}},\ and\ \bibinfo {author} {\bibfnamefont
  {D.}~\bibnamefont {Suess}},\ }\bibfield  {title} {\bibinfo {title}
  {{NeuralMag}: an open-source nodal finite-difference code for inverse
  micromagnetics},\ }\href {https://doi.org/10.1038/s41524-025-01688-1}
  {\bibfield  {journal} {\bibinfo  {journal} {npj Computational Materials}\
  }\textbf {\bibinfo {volume} {11}},\ \bibinfo {pages} {193} (\bibinfo {year}
  {2025})}\BibitemShut {NoStop}%
\bibitem [{\citenamefont {Bruckner}\ \emph {et~al.}(2023)\citenamefont
  {Bruckner}, \citenamefont {Koraltan}, \citenamefont {Abert},\ and\
  \citenamefont {Suess}}]{bruckner_magnumnp_2023}%
  \BibitemOpen
  \bibfield  {author} {\bibinfo {author} {\bibfnamefont {F.}~\bibnamefont
  {Bruckner}}, \bibinfo {author} {\bibfnamefont {S.}~\bibnamefont {Koraltan}},
  \bibinfo {author} {\bibfnamefont {C.}~\bibnamefont {Abert}},\ and\ \bibinfo
  {author} {\bibfnamefont {D.}~\bibnamefont {Suess}},\ }\bibfield  {title}
  {\bibinfo {title} {magnum.np: a {PyTorch} based {GPU} enhanced finite
  difference micromagnetic simulation framework for high level development and
  inverse design},\ }\href {https://doi.org/10.1038/s41598-023-39192-5}
  {\bibfield  {journal} {\bibinfo  {journal} {Scientific Reports}\ }\textbf
  {\bibinfo {volume} {13}},\ \bibinfo {pages} {12054} (\bibinfo {year}
  {2023})}\BibitemShut {NoStop}%
\bibitem [{sup()}]{supplemental_material}%
  \BibitemOpen
  \href@noop {} {}\bibinfo {note} {See Supplemental Material at the end of this
  document for the upward-continuation derivation, simulation setup,
  sign-ambiguity analysis, synthetic validation and diagnostics, sensitivity
  study, and the physical composition of the extracted effective distances,
  which includes Refs.~\cite{bodenberger_zur_1977, hubert_magnetic_2014,
  rohart_skyrmion_2013, bradbury_jax_2021, jaxopt_implicit_2022,
  kingma_adam_2017, becigneul2019riemannianadaptiveoptimizationmethods,
  kochurov_geoopt_2020, magnumnp_developers_inverse_2026, paszke_pytorch_2019,
  xie_air_2024}.}\BibitemShut {Stop}%
\bibitem [{\citenamefont {Abert}(2019)}]{abert_micromagnetics_2019}%
  \BibitemOpen
  \bibfield  {author} {\bibinfo {author} {\bibfnamefont {C.}~\bibnamefont
  {Abert}},\ }\bibfield  {title} {\bibinfo {title} {Micromagnetics and
  spintronics: models and numerical methods},\ }\href
  {https://doi.org/10.1140/epjb/e2019-90599-6} {\bibfield  {journal} {\bibinfo
  {journal} {The European Physical Journal B}\ }\textbf {\bibinfo {volume}
  {92}},\ \bibinfo {pages} {120} (\bibinfo {year} {2019})}\BibitemShut
  {NoStop}%
\bibitem [{\citenamefont {Zhang}\ \emph {et~al.}(2024)\citenamefont {Zhang},
  \citenamefont {Jiang}, \citenamefont {Jiang}, \citenamefont {He},
  \citenamefont {Zhang}, \citenamefont {Hu}, \citenamefont {Zhao},
  \citenamefont {Yang}, \citenamefont {Liu}, \citenamefont {Peng},
  \citenamefont {Yang},\ and\ \citenamefont
  {Yang}}]{zhang_above-room-temperature_2024}%
  \BibitemOpen
  \bibfield  {author} {\bibinfo {author} {\bibfnamefont {C.}~\bibnamefont
  {Zhang}}, \bibinfo {author} {\bibfnamefont {Z.}~\bibnamefont {Jiang}},
  \bibinfo {author} {\bibfnamefont {J.}~\bibnamefont {Jiang}}, \bibinfo
  {author} {\bibfnamefont {W.}~\bibnamefont {He}}, \bibinfo {author}
  {\bibfnamefont {J.}~\bibnamefont {Zhang}}, \bibinfo {author} {\bibfnamefont
  {F.}~\bibnamefont {Hu}}, \bibinfo {author} {\bibfnamefont {S.}~\bibnamefont
  {Zhao}}, \bibinfo {author} {\bibfnamefont {D.}~\bibnamefont {Yang}}, \bibinfo
  {author} {\bibfnamefont {Y.}~\bibnamefont {Liu}}, \bibinfo {author}
  {\bibfnamefont {Y.}~\bibnamefont {Peng}}, \bibinfo {author} {\bibfnamefont
  {H.}~\bibnamefont {Yang}},\ and\ \bibinfo {author} {\bibfnamefont
  {H.}~\bibnamefont {Yang}},\ }\bibfield  {title} {\bibinfo {title}
  {Above-room-temperature chiral skyrmion lattice and
  {Dzyaloshinskii}–{Moriya} interaction in a van der {Waals} ferromagnet
  {$Fe_{3-x}GaTe_2$}},\ }\href {https://doi.org/10.1038/s41467-024-48799-9}
  {\bibfield  {journal} {\bibinfo  {journal} {Nature Communications}\ }\textbf
  {\bibinfo {volume} {15}},\ \bibinfo {pages} {4472} (\bibinfo {year}
  {2024})}\BibitemShut {NoStop}%
\bibitem [{\citenamefont {Rondin}\ \emph {et~al.}(2014)\citenamefont {Rondin},
  \citenamefont {Tetienne}, \citenamefont {Hingant}, \citenamefont {Roch},
  \citenamefont {Maletinsky},\ and\ \citenamefont
  {Jacques}}]{rondin_magnetometry_2014}%
  \BibitemOpen
  \bibfield  {author} {\bibinfo {author} {\bibfnamefont {L.}~\bibnamefont
  {Rondin}}, \bibinfo {author} {\bibfnamefont {J.-P.}\ \bibnamefont
  {Tetienne}}, \bibinfo {author} {\bibfnamefont {T.}~\bibnamefont {Hingant}},
  \bibinfo {author} {\bibfnamefont {J.-F.}\ \bibnamefont {Roch}}, \bibinfo
  {author} {\bibfnamefont {P.}~\bibnamefont {Maletinsky}},\ and\ \bibinfo
  {author} {\bibfnamefont {V.}~\bibnamefont {Jacques}},\ }\bibfield  {title}
  {\bibinfo {title} {Magnetometry with nitrogen-vacancy defects in diamond},\
  }\href {https://doi.org/10.1088/0034-4885/77/5/056503} {\bibfield  {journal}
  {\bibinfo  {journal} {Reports on Progress in Physics}\ }\textbf {\bibinfo
  {volume} {77}},\ \bibinfo {pages} {056503} (\bibinfo {year}
  {2014})}\BibitemShut {NoStop}%
\bibitem [{\citenamefont {Blakely}(1995)}]{blakely_potential_1995}%
  \BibitemOpen
  \bibfield  {author} {\bibinfo {author} {\bibfnamefont {R.~J.}\ \bibnamefont
  {Blakely}},\ }\href@noop {} {\emph {\bibinfo {title} {Potential Theory in
  Gravity and Magnetic Applications}}}\ (\bibinfo  {publisher} {Cambridge
  University Press},\ \bibinfo {address} {Cambridge},\ \bibinfo {year}
  {1995})\BibitemShut {NoStop}%
\bibitem [{\citenamefont {Landau}\ and\ \citenamefont
  {Lifshitz}(1935)}]{landau_theory_1935}%
  \BibitemOpen
  \bibfield  {author} {\bibinfo {author} {\bibfnamefont {L.~D.}\ \bibnamefont
  {Landau}}\ and\ \bibinfo {author} {\bibfnamefont {E.~M.}\ \bibnamefont
  {Lifshitz}},\ }\bibfield  {title} {\bibinfo {title} {Theory of the dispersion
  of magnetic permeability in ferromagnetic bodies},\ }\href@noop {} {\bibfield
   {journal} {\bibinfo  {journal} {Physikalische Zeitschrift der Sowjetunion}\
  }\textbf {\bibinfo {volume} {8}},\ \bibinfo {pages} {153} (\bibinfo {year}
  {1935})}\BibitemShut {NoStop}%
\bibitem [{\citenamefont {Gilbert}(2004)}]{gilbert_phenomenological_2004}%
  \BibitemOpen
  \bibfield  {author} {\bibinfo {author} {\bibfnamefont {T.~L.}\ \bibnamefont
  {Gilbert}},\ }\bibfield  {title} {\bibinfo {title} {A phenomenological theory
  of damping in ferromagnetic materials},\ }\href@noop {} {\bibfield  {journal}
  {\bibinfo  {journal} {IEEE Transactions on Magnetics}\ }\textbf {\bibinfo
  {volume} {40}},\ \bibinfo {pages} {3443} (\bibinfo {year}
  {2004})}\BibitemShut {NoStop}%
\bibitem [{\citenamefont {Bodenberger}\ and\ \citenamefont
  {Hubert}(1977)}]{bodenberger_zur_1977}%
  \BibitemOpen
  \bibfield  {author} {\bibinfo {author} {\bibfnamefont {R.}~\bibnamefont
  {Bodenberger}}\ and\ \bibinfo {author} {\bibfnamefont {A.}~\bibnamefont
  {Hubert}},\ }\bibfield  {title} {\bibinfo {title} {Zur bestimmung der
  blochwandenergie von einachsigen ferromagneten},\ }\href
  {https://doi.org/10.1002/pssa.2210440146} {\bibfield  {journal} {\bibinfo
  {journal} {physica status solidi (a)}\ }\textbf {\bibinfo {volume} {44}},\
  \bibinfo {pages} {K7} (\bibinfo {year} {1977})},\ \bibinfo {note} {\_eprint:
  https://onlinelibrary.wiley.com/doi/pdf/10.1002/pssa.2210440146}\BibitemShut
  {NoStop}%
\bibitem [{\citenamefont {Hubert}\ and\ \citenamefont
  {Schäfer}(2014)}]{hubert_magnetic_2014}%
  \BibitemOpen
  \bibfield  {author} {\bibinfo {author} {\bibfnamefont {A.}~\bibnamefont
  {Hubert}}\ and\ \bibinfo {author} {\bibfnamefont {R.}~\bibnamefont
  {Schäfer}},\ }\href@noop {} {\emph {\bibinfo {title} {Magnetic domains: the
  analysis of magnetic microstructures}}},\ \bibinfo {edition} {softcover
  reprint of the hardcover 1st ed. 1998, corrected printing 2000}\ ed.\
  (\bibinfo  {publisher} {Springer},\ \bibinfo {address} {Berlin Heidelberg},\
  \bibinfo {year} {2014})\BibitemShut {NoStop}%
\bibitem [{\citenamefont {Rohart}\ and\ \citenamefont
  {Thiaville}(2013)}]{rohart_skyrmion_2013}%
  \BibitemOpen
  \bibfield  {author} {\bibinfo {author} {\bibfnamefont {S.}~\bibnamefont
  {Rohart}}\ and\ \bibinfo {author} {\bibfnamefont {A.}~\bibnamefont
  {Thiaville}},\ }\bibfield  {title} {\bibinfo {title} {Skyrmion confinement in
  ultrathin film nanostructures in the presence of {Dzyaloshinskii}-{Moriya}
  interaction},\ }\href {https://doi.org/10.1103/PhysRevB.88.184422} {\bibfield
   {journal} {\bibinfo  {journal} {Physical Review B}\ }\textbf {\bibinfo
  {volume} {88}},\ \bibinfo {pages} {184422} (\bibinfo {year}
  {2013})}\BibitemShut {NoStop}%
\bibitem [{\citenamefont {Bradbury}\ \emph {et~al.}(2021)\citenamefont
  {Bradbury}, \citenamefont {Frostig}, \citenamefont {Hawkins}, \citenamefont
  {Johnson}, \citenamefont {Leary}, \citenamefont {Maclaurin}, \citenamefont
  {Necula}, \citenamefont {Paszke}, \citenamefont {VanderPlas}, \citenamefont
  {Wanderman-Milne},\ and\ \citenamefont {Zhang}}]{bradbury_jax_2021}%
  \BibitemOpen
  \bibfield  {author} {\bibinfo {author} {\bibfnamefont {J.}~\bibnamefont
  {Bradbury}}, \bibinfo {author} {\bibfnamefont {R.}~\bibnamefont {Frostig}},
  \bibinfo {author} {\bibfnamefont {P.}~\bibnamefont {Hawkins}}, \bibinfo
  {author} {\bibfnamefont {M.~J.}\ \bibnamefont {Johnson}}, \bibinfo {author}
  {\bibfnamefont {C.}~\bibnamefont {Leary}}, \bibinfo {author} {\bibfnamefont
  {D.}~\bibnamefont {Maclaurin}}, \bibinfo {author} {\bibfnamefont
  {G.}~\bibnamefont {Necula}}, \bibinfo {author} {\bibfnamefont
  {A.}~\bibnamefont {Paszke}}, \bibinfo {author} {\bibfnamefont
  {J.}~\bibnamefont {VanderPlas}}, \bibinfo {author} {\bibfnamefont
  {S.}~\bibnamefont {Wanderman-Milne}},\ and\ \bibinfo {author} {\bibfnamefont
  {Q.}~\bibnamefont {Zhang}},\ }\bibfield  {title} {\bibinfo {title} {{JAX}:
  {Autograd} and {XLA}},\ }\href
  {https://ui.adsabs.harvard.edu/abs/2021ascl.soft11002B} {\bibfield  {journal}
  {\bibinfo  {journal} {Astrophysics Source Code Library}\ ,\ \bibinfo {pages}
  {ascl:2111.002}} (\bibinfo {year} {2021})},\ \bibinfo {note} {aDS Bibcode:
  2021ascl.soft11002B}\BibitemShut {NoStop}%
\bibitem [{\citenamefont {Blondel}\ \emph {et~al.}(2022)\citenamefont
  {Blondel}, \citenamefont {Berthet}, \citenamefont {Cuturi}, \citenamefont
  {Frostig}, \citenamefont {Hoyer}, \citenamefont {Llinares-L{\'o}pez},
  \citenamefont {Pedregosa},\ and\ \citenamefont
  {Vert}}]{jaxopt_implicit_2022}%
  \BibitemOpen
  \bibfield  {author} {\bibinfo {author} {\bibfnamefont {M.}~\bibnamefont
  {Blondel}}, \bibinfo {author} {\bibfnamefont {Q.}~\bibnamefont {Berthet}},
  \bibinfo {author} {\bibfnamefont {M.}~\bibnamefont {Cuturi}}, \bibinfo
  {author} {\bibfnamefont {R.}~\bibnamefont {Frostig}}, \bibinfo {author}
  {\bibfnamefont {S.}~\bibnamefont {Hoyer}}, \bibinfo {author} {\bibfnamefont
  {F.}~\bibnamefont {Llinares-L{\'o}pez}}, \bibinfo {author} {\bibfnamefont
  {F.}~\bibnamefont {Pedregosa}},\ and\ \bibinfo {author} {\bibfnamefont
  {J.-P.}\ \bibnamefont {Vert}},\ }\bibfield  {title} {\bibinfo {title}
  {Efficient and modular implicit differentiation},\ }in\ \href@noop {} {\emph
  {\bibinfo {booktitle} {Advances in Neural Information Processing Systems}}},\
  Vol.~\bibinfo {volume} {35}\ (\bibinfo {year} {2022})\ pp.\ \bibinfo {pages}
  {5230--5242}\BibitemShut {NoStop}%
\bibitem [{\citenamefont {Kingma}\ and\ \citenamefont
  {Ba}(2017)}]{kingma_adam_2017}%
  \BibitemOpen
  \bibfield  {author} {\bibinfo {author} {\bibfnamefont {D.~P.}\ \bibnamefont
  {Kingma}}\ and\ \bibinfo {author} {\bibfnamefont {J.}~\bibnamefont {Ba}},\
  }\href {https://doi.org/10.48550/arXiv.1412.6980} {\bibinfo {title} {Adam:
  {A} {Method} for {Stochastic} {Optimization}}} (\bibinfo {year} {2017}),\
  \bibinfo {note} {arXiv:1412.6980 [cs]}\BibitemShut {NoStop}%
\bibitem [{\citenamefont {Bécigneul}\ and\ \citenamefont
  {Ganea}(2019)}]{becigneul2019riemannianadaptiveoptimizationmethods}%
  \BibitemOpen
  \bibfield  {author} {\bibinfo {author} {\bibfnamefont {G.}~\bibnamefont
  {Bécigneul}}\ and\ \bibinfo {author} {\bibfnamefont {O.-E.}\ \bibnamefont
  {Ganea}},\ }\href {https://arxiv.org/abs/1810.00760} {\bibinfo {title}
  {Riemannian adaptive optimization methods}} (\bibinfo {year} {2019}),\
  \Eprint {https://arxiv.org/abs/1810.00760} {arXiv:1810.00760 [cs.LG]}
  \BibitemShut {NoStop}%
\bibitem [{\citenamefont {Kochurov}\ \emph {et~al.}(2020)\citenamefont
  {Kochurov}, \citenamefont {Karimov},\ and\ \citenamefont
  {Kozlukov}}]{kochurov_geoopt_2020}%
  \BibitemOpen
  \bibfield  {author} {\bibinfo {author} {\bibfnamefont {M.}~\bibnamefont
  {Kochurov}}, \bibinfo {author} {\bibfnamefont {R.}~\bibnamefont {Karimov}},\
  and\ \bibinfo {author} {\bibfnamefont {S.}~\bibnamefont {Kozlukov}},\ }\href
  {https://doi.org/10.48550/arXiv.2005.02819} {\bibinfo {title} {Geoopt:
  {Riemannian} {Optimization} in {PyTorch}}} (\bibinfo {year} {2020}),\
  \bibinfo {note} {arXiv:2005.02819 [cs]}\BibitemShut {NoStop}%
\bibitem [{\citenamefont {{magnum.np
  developers}}(2026)}]{magnumnp_developers_inverse_2026}%
  \BibitemOpen
  \bibfield  {author} {\bibinfo {author} {\bibnamefont {{magnum.np
  developers}}},\ }\href@noop {} {\bibinfo {title} {{Inverse Problems} ---
  {magnum.np} 2.2.0 documentation}},\ \bibinfo {howpublished}
  {\url{https://magnum.np.gitlab.io/magnum.np/inverse_problems.html}} (\bibinfo
  {year} {2026}),\ \bibinfo {note} {accessed: 2026-02-18}\BibitemShut {NoStop}%
\bibitem [{\citenamefont {Paszke}\ \emph {et~al.}(2019)\citenamefont {Paszke},
  \citenamefont {Gross}, \citenamefont {Massa}, \citenamefont {Lerer},
  \citenamefont {Bradbury}, \citenamefont {Chanan}, \citenamefont {Killeen},
  \citenamefont {Lin}, \citenamefont {Gimelshein}, \citenamefont {Antiga},
  \citenamefont {Desmaison}, \citenamefont {Köpf}, \citenamefont {Yang},
  \citenamefont {DeVito}, \citenamefont {Raison}, \citenamefont {Tejani},
  \citenamefont {Chilamkurthy}, \citenamefont {Steiner}, \citenamefont {Fang},\
  and\ \citenamefont {Chintala}}]{paszke_pytorch_2019}%
  \BibitemOpen
  \bibfield  {author} {\bibinfo {author} {\bibfnamefont {A.}~\bibnamefont
  {Paszke}}, \bibinfo {author} {\bibfnamefont {S.}~\bibnamefont {Gross}},
  \bibinfo {author} {\bibfnamefont {F.}~\bibnamefont {Massa}}, \bibinfo
  {author} {\bibfnamefont {A.}~\bibnamefont {Lerer}}, \bibinfo {author}
  {\bibfnamefont {J.}~\bibnamefont {Bradbury}}, \bibinfo {author}
  {\bibfnamefont {G.}~\bibnamefont {Chanan}}, \bibinfo {author} {\bibfnamefont
  {T.}~\bibnamefont {Killeen}}, \bibinfo {author} {\bibfnamefont
  {Z.}~\bibnamefont {Lin}}, \bibinfo {author} {\bibfnamefont {N.}~\bibnamefont
  {Gimelshein}}, \bibinfo {author} {\bibfnamefont {L.}~\bibnamefont {Antiga}},
  \bibinfo {author} {\bibfnamefont {A.}~\bibnamefont {Desmaison}}, \bibinfo
  {author} {\bibfnamefont {A.}~\bibnamefont {Köpf}}, \bibinfo {author}
  {\bibfnamefont {E.}~\bibnamefont {Yang}}, \bibinfo {author} {\bibfnamefont
  {Z.}~\bibnamefont {DeVito}}, \bibinfo {author} {\bibfnamefont
  {M.}~\bibnamefont {Raison}}, \bibinfo {author} {\bibfnamefont
  {A.}~\bibnamefont {Tejani}}, \bibinfo {author} {\bibfnamefont
  {S.}~\bibnamefont {Chilamkurthy}}, \bibinfo {author} {\bibfnamefont
  {B.}~\bibnamefont {Steiner}}, \bibinfo {author} {\bibfnamefont
  {L.}~\bibnamefont {Fang}},\ and\ \bibinfo {author} {\bibfnamefont
  {S.}~\bibnamefont {Chintala}},\ }\href
  {https://doi.org/10.48550/arXiv.1912.01703} {\emph {\bibinfo {title}
  {{PyTorch}: {An} {Imperative} {Style}, {High}-{Performance} {Deep} {Learning}
  {Library}}}}\ (\bibinfo {year} {2019})\BibitemShut {NoStop}%
\bibitem [{\citenamefont {Xie}\ \emph {et~al.}(2024)\citenamefont {Xie},
  \citenamefont {Zhang}, \citenamefont {Bai}, \citenamefont {Liu},
  \citenamefont {Wang}, \citenamefont {Yu}, \citenamefont {Li}, \citenamefont
  {Chang}, \citenamefont {Wang}, \citenamefont {Gao}, \citenamefont {Wei},
  \citenamefont {Zhao},\ and\ \citenamefont {Nie}}]{xie_air_2024}%
  \BibitemOpen
  \bibfield  {author} {\bibinfo {author} {\bibfnamefont {W.}~\bibnamefont
  {Xie}}, \bibinfo {author} {\bibfnamefont {J.}~\bibnamefont {Zhang}}, \bibinfo
  {author} {\bibfnamefont {Y.}~\bibnamefont {Bai}}, \bibinfo {author}
  {\bibfnamefont {Y.}~\bibnamefont {Liu}}, \bibinfo {author} {\bibfnamefont
  {H.}~\bibnamefont {Wang}}, \bibinfo {author} {\bibfnamefont {P.}~\bibnamefont
  {Yu}}, \bibinfo {author} {\bibfnamefont {J.}~\bibnamefont {Li}}, \bibinfo
  {author} {\bibfnamefont {H.}~\bibnamefont {Chang}}, \bibinfo {author}
  {\bibfnamefont {Z.}~\bibnamefont {Wang}}, \bibinfo {author} {\bibfnamefont
  {F.}~\bibnamefont {Gao}}, \bibinfo {author} {\bibfnamefont {G.}~\bibnamefont
  {Wei}}, \bibinfo {author} {\bibfnamefont {W.}~\bibnamefont {Zhao}},\ and\
  \bibinfo {author} {\bibfnamefont {T.}~\bibnamefont {Nie}},\ }\bibfield
  {title} {\bibinfo {title} {Air stability and composition evolution in van der
  {Waals} {$Fe_3GeTe_2$}},\ }\href {https://doi.org/10.1063/5.0194520}
  {\bibfield  {journal} {\bibinfo  {journal} {APL Materials}\ }\textbf
  {\bibinfo {volume} {12}},\ \bibinfo {pages} {031102} (\bibinfo {year}
  {2024})}\BibitemShut {NoStop}%
\end{thebibliography}%
